\documentclass[a4paper,twoside,english]{revtex4}
\usepackage[T1]{fontenc}
\usepackage[latin1]{inputenc}
\usepackage{verbatim}
\usepackage{amsmath}
\usepackage{color}
\usepackage{amssymb}

\makeatletter

\providecommand{\tabularnewline}{\\}

\usepackage{babel}
\makeatother
\begin{document}

\title{Birational Mappings and Matrix Sub-algebra from the Chiral Potts
Model}

\author{E. Preissmann$^{\star}$, J.-Ch. Anglès d'Auriac$^{\dagger}$, J.-M.
Maillard$^{\ddagger}$}

\address{$^{\star}$Emmanuel.Preissmann@gmail.com}

\address{$^{\dagger}$dauriac@grenoble.cnrs.fr}

\address{$^{\ddagger}$maillard@lptmc.jussieu.fr}

\affiliation{$^{\dagger}$Institut Neel département MCBT 25 Av. des martyrs BP
166 38042 Grenoble Cedex 09 France. }

\affiliation{$^{\ddagger}$LPTMC, Tour 24, case 121, 4, Place Jussieu, 75252 Paris
Cedex 05, France. }

\date{July $11{}^{\mbox{th}}$, 2008}

\begin{abstract}
We study birational transformations of the projective space originating
from lattice statistical mechanics, specifically from various chiral
Potts models. Associating these models to \emph{stable patterns} and
\emph{signed-patterns}, we give general results which allow us to
find \emph{all} chiral $q$-state spin-edge Potts models when the
number of states $q$ is a prime or the square of a prime, as well
as several $q$-dependent family of models. We also prove the absence
of monocolor stable signed-pattern with more than four states. This
demonstrates a conjecture about cyclic Hadamard matrices in a particular
case. The birational transformations associated to these lattice spin-edge
models show complexity reduction. In particular we recover a one-parameter
family of integrable transformations, for which we give a matrix representation
when the parameter has a suitable value. 
\end{abstract}
\maketitle

\section{Introduction and presentation}

In a previous publication \cite{OnTheComp} a set of birational transformations
acting on projective spaces of various dimension has been introduced.
These transformations are birational realizations of Coxeter groups.
They arise naturally in lattice statistical mechanics in relation
with the Yang-Baxter equations for solving vertex models and star-triangle
relation for solving spin model \cite{biratfrommecastat,BaBook}.
However it is important to note that these birational symmetries are
actually symmetries of the lattice of statistical mechanics models
beyond the Yang-Baxter integrable situations: they can be seen as
(generically infinite) discrete and non-linear symmetries of the parameter
space of the model and, for instance, of the phase diagram of these
lattice models \cite{MeAnMaRo94}. These transformations can in fact
be considered \emph{per se}, as discrete dynamical system. The degree
complexity (or entropy) of these transformations has been evaluated
\cite{definingcomplex1,definingcomplex2,definingcomplex3}. An unexpected
complexity reduction has been found, and it has been conjectured that
the most general Potts model has the \emph{same} complexity as the
most general cyclic Potts model. Considering that the most general
cyclic Chiral Potts model has only $q$ homogeneous parameters while
the most general Potts model has $q^{2}$ homogeneous parameters the
equality between their respective complexities is not obvious. In
this paper we go further and study many particular cyclic spin-edge
Potts Models and their associated birational transformations. As it
is described below finding spin-edge cyclic chiral Potts Models for
lattice lattice statistical mechanics amounts to finding the so-called
stable patterns. This problem turns out to be related to many interesting
field of mathematics: Bose-Mesner algebra \cite{BoseMesner}, association
schemes \cite{asssch_jaeger}, Hadamard matrices (one of our result
demonstrates a particular case of a conjecture about Hadamard matrices
\cite{conjecthada,CyclotomicInteger}), Gauss identity, etc .

A $q$-state Potts model \cite{FWu} is completely defined by a lattice
and a Boltzmann weight matrix $W$. The spins, which have $q$ states,
are located at the vertices of a graph, with \emph{oriented} edges.
The Boltzmann weight of a given spin configuration is the product
of the Boltzmann weight over all edges, hence the name \emph{spin-edge}
model. The Boltzmann weight of the edges can be conveniently seen
as a matrix : the rows refer to the beginning $i$ of the oriented
edge, and the column to the end $j$. The weight of the oriented edge
$(i,j)$ is $W(\sigma_{i},\sigma_{j})$. The so-called inverse-relation
\cite{inverserelation,inversrelatJM} implies a functional relation
between the partition function of a model associated to a matrix and
the model associated to its inverse for the same lattice. 

By definition a chiral Potts model is a model for which the entries
$W_{ij}$ and $W_{ji}$ are different, \emph{i.e.} the Boltzmann weight
matrix is not symmetric. Of particular interest are the cyclic chiral
models for which the Boltzmann weights $W_{ij}$ are functions of
$i-j\ \mbox{mod}\ q$. This class contains in particular the integrable
chiral Potts models \cite{chiralpotts1,chiralpotts2}. The global
symmetries of the cyclic models have been classified in \cite{MarcuRittenberg}
. The most general cyclic Potts model correspond to the case where
there is no other constraint. It means that the Boltzmann weight matrix
$W$ is cyclic. Now we can look for other less general models, obtained
by imposing further constraints on the entries of $W$. The simplest
constraints are equalities between some entries of the matrix $W$,
but we will also consider the case where these constraints are {}``anti-equalities'',
i.e. we demand that some pairs of entries are opposite. Imposing that
two Boltzmann weights are opposite could appear unphysical but, as
it will be explained below, it is mathematically very natural. 

However these constraints need to be compatible with the inversion
relation mentioned above, as well as with a Hadamard inverse described
below. The aim of this paper is to find such matrices and the associated
birational transformations. It is organized as follows: we first recall
some definitions and what is already known on this problem. We then
generalize the notions used in this framework and gather together
the notations we use. The next two sections are devoted to our analytical
results. These results are grouped into two sections, depending whether
the result is directly of importance from the lattice statistical
mechanics point of view, or not. In section\ref{sec:Analytical-Results-math}
we first present our more mathematical results. In contrast, the results
corresponding to lattice statistical physics are all given in the
subsection \ref{sub:List-of-theB} while the proofs, together with
comments and examples, are given in subsection \ref{sub:Proof-or-justificationB}.
The content of the next section \ref{sec:Analytical-Results-Meca-stat}
is more {}``Potts model oriented'' and as far as (Boltzmann weight
spin-edge) matrices are concerned, focused on particular subcases
of cyclic matrices, with an attempt to perform an exhaustive classification
of the \char`\"{}interesting spin-edge Potts models''. We give \emph{all}
chiral Potts models when the number of colors $q$ is a square or
the square of a prime. In some cases we also study the degree-complexity
of the birational transformations canonically associated with these
subcases of cyclic matrices.  Here also results are gathered in the
subsection\ref{sub:List-of-theB} and the proofs in subsection \ref{sub:Proof-or-justificationB}.
Then, in section \ref{sec:Numerical-results}, we turn to the cases
where we were not able to find analytical results and introduce a
computer-aided method which enable us to perform some calculation
despite the huge combinatorial of this problem. We then present these
results. 

As far as applications to spin-edge $q$-state Potts models are concerned
our main results correspond to finding the stable patterns, or in
other words, the \char`\"{}interesting lattice statistical mechanics
spin-edge models\char`\"{}, when, $q$, the number of states of the
$q$-state Potts model is a prime, or the square of a prime, and providing
some first steps for results when the number of states is the product
of two primes. We also have, as a byproduct, other more specific results,
like a demonstration of a conjecture about Hadamard matrices (for
monocolor stable patterns). Beside the rigorous proofs, we give numerous
examples, most of them in the appendices. We have tried to be rigorous
in the demonstration but pedagogical in the examples. The reader more
interested in the {}``lattice statistical mechanics point of view''
can skip the more mathematical section \ref{sec:Analytical-Results-math},
in particular the lemma which are more technical. However we feel
that the results of this section are worthwhile \emph{per se}, and
can be usefull to go further in the classification of the Potts models.

\subsection{Recalls}

\subsubsection{The context}

Starting from the lattice statistical mechanics point of view, we
consider an anisotropic Potts model on a square lattice with Boltzmann
weight matrix $W_{h}$ for the horizontal edges and $W_{v}$ for the
vertical edges. It has been shown \cite{inverserelation} that if
$T(W_{h},W_{v})$ is the transfer matrix of this model then \[
T(W_{h},W_{v})T(W_{h}^{-1},J(W_{v}))=C(W_{v})\mathbb{I}\]
where $J(W)$ designates the matrix which entries are the inverse
of the entries of $W$ (see below). Transporting this equality to
the eigenvalues of $T$ permits to find a functional relation for
the partition function of the model. This functional relation induces
a constraining symmetry for the phase diagram.

We now adopt a more general point of view and we consider the $q\times q$
matrices projectively as elements of $\mathbb{CP}_{q^{2}-1}$ (since
Boltzmann weight matrices are defined up to a multiplicative constant).
Using the same notation as in ref \cite{ClassifAMV}, we define $K=I\circ J$
where $I$ is the usual matrix inverse $I(M)=M^{-1}$ and $J$ is
the Hadamard inverse (inverse of the Hadamard product) defined by
$\left(J(M)\right)_{ij}=\frac{1}{M_{ij}}$. The transformations $I$
and $J$ are two non commuting involutions, which can be represented
polynomially in $\mathbb{CP}_{q^{2}-1}$. In this representation $I$
replaces each entry of $M$ by its cofactor, and $J$ replaces each
entry by the product of all other entries. It is clear that $K$ and
its inverse $K^{-1}=J\circ I$ are both rational transformations.
At each step, the $q^{2}$ entries of the matrix $M$ are factorized
as products of polynomial with integer coefficients, and the common
factors of all the entries are discarded.

\subsubsection{Degree complexity}

A quantity characterizing the complexity is the degree complexity
$\lambda$ \cite{definingcomplex1,definingcomplex2,definingcomplex3}.
We simply recall the definition\[
\lambda=\lim_{n\rightarrow\infty}\frac{1}{n}\log d_{n}\]
where $d_{n}$ is the degree of the $n^{\textrm{th }}$iterate $K^{n}$,
where $K$ is represented as $q$ homogeneous polynomials of degree
$d$. Without the factorizations $d_{n}=d^{n}$ and consequently $\lambda=\log d$.
For some transformations one has $d_{n}\sim\delta^{n}$ with $\delta<d$,
this is called a \emph{complexity reduction} \cite{bedfordkim,complexitereduc}.
When the growth of the degree is polynomial one has $\lambda=0$ and
the transformation is integrable\cite{moreintegmap}. Finally we define
the degree generating function as\begin{equation}
f(x)=\sum_{n=0}^{\infty}d_{n}x^{n}\label{eq:genfun}\end{equation}
when this serie has a positive radius of convergence.

\subsubsection{(I,J)-stable patterns }

We consider a disjoint partition $P=\left\{ E_{0},\cdots,E_{r-1}\right\} $
of the indices with $\bigsqcup_{k=0}^{r-1}E_{k}=\left\{ (i,j),\ i,j=0,\cdots,n-1\right\} $
where the symbol $\bigsqcup$ denotes the disjoint union, we call
$r$ the number of colors, and we consider a matrix $M$ such that
\[
(i,j)\in E_{k}\textrm{ and }(i',j')\in E_{k}\Rightarrow M_{ij}=M_{i'j'}\]
A matrix verifying this set of equalities is said to belong to the
\emph{pattern} $P$ as in ref \cite{OnTheComp}. We are interested
in the matrices belonging to the same pattern as their image by $K$.
Therefore we will consider pattern containing at least one invertible
matrix (for example we exclude the pattern where all entries are equal).
Obviously the transformation $J$ is compatible with any pattern.
Therefore a matrix and its $K-$image will belong to the same pattern
\emph{iff} this matrix and its inverse belong to the same pattern.
Such a pattern is said inverse-stable. The number $r$ of subset of
the partition is called the number of colors. The transformation associated
with a stable pattern with $r$ colors acts on $\mathbb{CP}_{r-1}$.

\subsubsection{Cyclic matrices}

Consider the set of $q\times q$ cyclic matrix $M_{c}$ with entries
$M_{c}(i,j)$ such that \begin{eqnarray}
M_{c}(i,j) & = & M_{c}\left(0,i-j\ \mbox{ mod }q\right)\label{eq:defcyclic}\end{eqnarray}
The corresponding model of lattice statistical mechanics is the cyclic
chiral Potts model\cite{chiralpotts1,chiralpotts2}. Chiral refers
to the fact that $M_{c}(i,j)$ is not necessarily equal to $M_{c}(j,i)$
(the lattice has an orientation) and cyclic refers to the fact that
one restricts to cyclic Boltzmann weight matrices. The corresponding
pattern is inverse-stable and also matrix-product stable (see next
sub section). Since a cyclic matrix is fully determined by its first
row, the transformation $K$ can be represented in $\mathbb{CP}_{q-1}$.
It is found that complexity reduction does occur for cyclic matrices
and the complexity is the largest root of $x^{2}+(2-(q-2)^{2})x+1$
\cite{bellonviallet}. From numerical analysis it has been conjectured
that this value for the algebraic complexity is the same than for
arbitrary matrices without any constraint on the entries. Another
inverse-stable pattern is provided by cyclic and symmetric matrices.
In that case the complexity reduction is even bigger and the complexity
is the root of largest modulus of $x^{2}+\left(2-(p-1)^{2}\right)x+1$
where $p=\left\lfloor \frac{q}{2}\right\rfloor +1$ where $\left\lfloor \right\rfloor $
denotes the integer part. One aim of this paper is to find some subspaces
where further complexity reduction takes place.

\subsection{Generalizations}

\subsubsection{Product-stability}

A pattern is said product-stable if the product of two matrices belonging
to this pattern also belongs to this pattern. Using the Cayley-Hamilton
theorem, one can express the inverse of a matrix $M$ as a linear
combination of its $q-1$ first powers. Therefore product-stability
implies inverse-stability. We are going to present examples where
the reciprocal proposition is wrong. From now on we call \emph{$P$-stable}
a pattern which is product-stable\emph{, $I$-stable} a pattern which
is inverse-stable, and $I\bar{P}$-stable a pattern inverse-stable
but not product-stable. An obvious example are the symmetric matrices
which are inverse stable (the inverse of a symmetric matrix is symmetric)
but are not product-stable (the matrix-product of two symmetric matrices
is not necessarily symmetric).

\subsubsection{Generalization of the notion of pattern : signed-patterns}

We generalize the notion of pattern and look for a set of $r$ independent
$q\times q$ matrices $M_{i}$ such that\[
K\left(\sum_{i=0}^{r-1}x_{i}M_{i}\right)=\sum_{i=0}^{r-1}y_{i}M_{i}\]

Let us introduce the characteristic function $\chi$ which associates
to each set of indices $E$ the matrix $\chi(E)$ defined by $\left(\chi(E)\right)_{ij}=\left\{ \begin{array}{ccc}
1 &  & (i,j)\in E\\
0 &  & (i,j)\notin E\end{array}\right.$. The patterns defined in the previous paragraph correspond to \begin{equation}
M_{k}=\chi(E_{k})\label{eq:patt}\end{equation}
For reasons explained below, we also consider the more general case
of matrices with entries 0, 1 or -1 \begin{equation}
M_{k}=\chi(E_{k}^{+})-\chi(E_{k}^{-})\label{eq:signepatt}\end{equation}
We call the partition $\left\{ E_{0}^{+},E_{0}^{-},\cdots,E_{r-1}^{+},E_{r-1}^{-}\right\} $
a \emph{signed-pattern}, and $r$ the number of colors. The algebra
generated by these matrices is $J$-stable. The notion of \char`\"{}signed-patterns\char`\"{}
simply corresponds to the notion of patterns defined by equalities
between entries \emph{up to a sign}.

\subsubsection{Stability of signed-patterns}

A product-stable set of matrices with entries 0 or 1 and which sum
up to the all-one entry matrix is called an \emph{association scheme}
\cite{asssch_jaeger}. It is an algebra. If the matrices are also
symmetric, then it is a \emph{Bose-Mesner} algebra \cite{BoseMesner}. 

The problem we address in this paper can be summarized as finding
$r$ $q\times q$ matrices $M_{i}$ with entry 0,1,-1 verifying\[
\sum_{i=1}^{r}\left|\left(M_{i}\right)_{jk}\right|=1\quad\forall j,k\]
and 

\begin{equation}
\left(\sum_{i=1}^{r}x_{i}M_{i}\right)\left(\sum_{i=1}^{r}y_{i}M_{i}\right)=\sum_{i=1}^{r}z_{i}M_{i}\label{eq:prod}\end{equation}
for $P$-stability and\begin{equation}
\left(\sum_{i=1}^{r}x_{i}M_{i}\right)^{-1}=\sum_{i=1}^{r}z_{i}M_{i}\label{eq:inv}\end{equation}
when the inverse of $\sum_{i=1}^{r}x_{i}M_{i}$ exists for $I$-stability.
From the definition, the matrices induced by a $P$-stable pattern
form an algebra. But the matrices induced by $I$-stable patterns
do not always (in contrast with the problem studied in \cite{MarcuRittenberg}). 

Note that if a set of matrices $\left\{ M_{i}\right\} $ defines an
algebra, so does the set $\left\{ P_{\sigma}^{-1}M_{i}P_{\sigma}\right\} $
where $\sigma$ is a permutation of $\left\{ 0,\cdots,q-1\right\} $
and $P$ the associated permutation matrix $\left(P_{\sigma}\right)_{ij}=\delta_{i,\sigma(j)}$.
However if $M$ is a cyclic matrix, $P_{\sigma}^{-1}MP_{\sigma}$
is not necessarily a cyclic matrix.

\subsection{\textcolor{black}{Notations and definitions}}

\begin{enumerate}
\item From now on we will restrict ourself to cyclic matrices. As far as
notations are concerned, we identify a cyclic matrix and its first
row seen as a vector in $\mathbb{CP}_{q-1}$. Let $\mathbf{v}\in\mathbb{C}^{q}$
be a vector, we use the notation $\mbox{Cy}(\mathbf{v})$ to denote
the $q\times q$ matrix \[
\left(\mbox{Cy}(\mathbf{v})\right)_{ij}=v_{i-j}\]
and $\mbox{Diag}(\mathbf{v})$ to denote the $q\times q$ matrix \[
\left(\mbox{Diag}(\mathbf{v})\right)_{ij}=v_{i}\delta_{ij}\]
$\delta$ is a Kronecker symbol. 
\item As already mentioned in the introduction, the discrete Fourier transform
plays a crucial role for stability of cyclic matrices. We therefore
define the matrix $U=\left(\omega^{ij}\right)$ where $\omega=\exp\frac{2\pi}{q}\imath$
and use notation\[
\widehat{\mathbf{x}}=U\mathbf{x}\]
to denote the Fourier transform of the vector $\mathbf{x}\in\mathbb{C}^{q}$.
We note the relation \begin{equation}
U^{\star}\times\mbox{Cy}(\mathbf{x})\times U=q\ \mbox{Diag}(\widehat{\mathbf{x}})\label{eq:diag}\end{equation}
which will be useful. For the reader familiar to lattice statistical
mechanics $U$ is the generalization of the Kramers-Wannier duality,
however it is \emph{not} a transformation of order two, but of order
four. 
\item When patterns are \emph{explicitly} given we use a straightforward
representation: we put in a bracket the entries of the first row.
An example is given with the comments in appendix \ref{app:q8}.
\item The subspace spanned by the set of vectors $\left\{ \mathbf{v}(i)\right\} _{i=1\cdots r}$
with complex coefficients is denoted: \[
\bigoplus_{1\le i\le r}\mathbb{C}\mathbf{v}(i)\]

\item We use arithmetic modulo $q$, $\mathbb{Z}_{q}^{\star}$ is the set
of elements of $\mathbb{Z}_{q}$ which are invertible and if $d$
is a divisor of $q$ (noted $d\mid q$) one introduces\[
\mathbb{Z}(q,d)=\left\{ k\in\mathbb{Z}_{q}\ \mid\mbox{gcd}(k,q)=d\right\} \]

\item \textcolor{black}{We will also use the convolution product (noted
$\star$) and the Hadamard product (noted $.$) between two vectors
of $\mathbb{C}^{q}$ defined respectively by\begin{eqnarray*}
\left(\mathbf{u}\star\mathbf{v}\right)_{i} & = & \sum_{j=0}^{q-1}u_{j}v_{i-j}\\
\left(\mathbf{u\cdot v}\right)_{i} & = & u_{i}v_{i}\end{eqnarray*}
It is straightforward to see that $\mbox{Cy}(\mathbf{u})\mbox{Cy}(\mathbf{v})=\mbox{Cy}(\mathbf{u}\star\mathbf{v})$.
With these notations, Eq. \ref{eq:diag} reads: \[
\widehat{\mathbf{u}\star\mathbf{v}}=q\widehat{\mathbf{u}}\cdot\widehat{\mathbf{v}}\]
We note $\mathbf{u}^{\star n}$ the convolution product of $\mathbf{u}$
with itself $n$ times. By convention $\mathbf{u}^{\star0}=\chi\left(\left\{ 0\right\} \right)$.
Keeping in mind that diagonalization of the cyclic matrices requires
the discrete Fourier transform Eq. \ref{eq:diag}, the previous relation
amounts to writing that the eigenvalues of the product of two cyclic
matrices is the product of the eigenvalues. With these notations a
partition $\mathcal{E}=\left\{ E_{1},\cdots,E_{r}\right\} $ is $P$-stable
if $\forall a_{i},b_{i},\ \exists c_{i}$ such that\[
\sum_{i}a_{i}\chi(E_{i})\star\sum_{i}b_{i}\chi(E_{i})=\sum_{i}c_{i}\chi(E_{i})\]
the partition $\mathcal{E}=\left\{ E_{1},\cdots,E_{r}\right\} $is
$I$-stable if $\forall a_{i},\ \exists c_{i}$ such that\[
\sum_{i}a_{i}\chi(E_{i})\star\sum_{i}c_{i}\chi(E_{i})=(1,0,\cdots,0)\]
corresponding to the matrix inversion of a cyclic matrix. }
\item \textcolor{black}{A set of disjoint subsets $\mathcal{E}=\left\{ E_{0}=\left\{ 0\right\} {,E}_{1},\cdots,E_{k}\right\} $
of $\left\{ 0,\cdots,q-1\right\} $ is} \textcolor{black}{\emph{convenient}}
\textcolor{black}{if $\forall\left(n_{1},\cdots,n_{k}\right)\in\mathbb{N}^{k}\ \forall l\in\left[0,k\right]$
\[
\forall i,j\in E_{l}\quad\left(\chi(E_{1})^{\star n_{1}}\star\cdots\star\chi(E_{k})^{\star n_{k}}\right)_{i}=\left(\chi(E_{1})^{\star n_{1}}\star\cdots\star\chi(E_{k})^{\star n_{k}}\right)_{j}\]
By a slight abuse of notation a set $E$ such that $\left\{ \left\{ 0\right\} ,E\right\} $
is convenient is also called convenient. In that case one has\begin{equation}
\forall n>0\quad\forall i,j\in E\quad\left(\mbox{Cy}(\chi(E))^{n}\right)_{0,i}=\left(\mbox{Cy}(\chi(E))^{n}\right)_{0,j}\label{eq:convenient}\end{equation}
Actually, since any power of a $q\times q$ $M$ can be expressed
as linear combination of the first $q-1$ powers with the help of
the Cayley-Hamilton theorem, one needs to verify Eq.\ref{eq:convenient}
only for $0<n\le q-1$. }\\
\textcolor{black}{Intuitively, a convenient set of disjoint subsets
can be seen as a possible {}``beginning'' of a stable pattern. Indeed
it verifies some} \textcolor{black}{\emph{necessary}} \textcolor{black}{conditions
such that it can be it extended to a stable pattern. In particular
each set of a stable partition is convenient. This will be used in
section \ref{sec:Numerical-results}. Note that if $\mathcal{E}$
is a partition then it is $P$-stable.}
\item \textcolor{black}{An} \textcolor{black}{\emph{admissible}} \textcolor{black}{set
is a subse}\textcolor{black}{\emph{t}} \textcolor{black}{$E$ of $\left\{ 0,\cdots,q-1\right\} $\begin{equation}
E=\bigsqcup_{d\in D}d\ i_{d}H_{d}\label{eq:Admissi}\end{equation}
where $D$ is a subset of the divisors of $q$, $H_{d}$ is a subgroup
of $\mathbb{Z}_{\frac{q}{d}}^{\star}$ and $\mbox{gcd}(di_{d},q)=d\;\forall d\in D$.
We will show below that the union in Eq. \ref{eq:Admissi} is indeed
a disjoint union and that the admissible sets are in fact the} \textcolor{black}{\emph{only}}
\textcolor{black}{possible sets in a stable pattern.}
\item For $I$ a finite set, $\mathbf{x}=(x_{i})_{i\in I}$ we note $\mathcal{P}_{\mathbf{x},I}=\left\{ A_{1},\cdots,A_{r}\right\} $
the partition of $I$ such that $x_{i}=x_{j}$ \emph{iff} $i$ and
$j$ are in a same $A_{k}$. Actually the $A_{k}$'s are the preimages
of the application $i\rightarrow x_{i}$. When the set $I=\left\{ 1,\cdots,q\right\} $
we do not specify it and note simply $\mathcal{P}_{\mathbf{x}}$.
For example for $q=4$ $\mathcal{P}_{(x_{1},x_{2},x_{1},x_{1})}=\left\{ \left\{ 1,3,4\right\} ,\left\{ 2\right\} \right\} $
\item Finally if $E$ is a group, $F<E$ means that $F$ is a subgroup of
$E$.
\end{enumerate}

\section{Analytical results pertaining to mathematics \label{sec:Analytical-Results-math}}

In this paragraph we express the inverse-stability and the product-stability
for the pattern and the signed-pattern in term of matrix subalgebra
and list mathematical results which are used in the next section.
We also present results interesting \emph{per se}, and likely to be
usefull to go further in the classification of the lattice models.

\subsection{List of the results\label{sub:List-of-theA}}

\subsubsection{Pattern stability as matrix subalgebra\label{sub:Patt-Subalg_R1}}

\begin{itemize}
\item The pattern $\mathcal{E}=\left\{ E_{i}\right\} $ is product-stable
\emph{iff} there exists a partition $\mathcal{F}=\left\{ F_{j}\right\} $
such that \begin{eqnarray}
\bigoplus_{1\le i\le r}\mathbb{C}\widehat{\chi(E_{i})} & = & \bigoplus_{1\le j\le r}\mathbb{C}\chi(F_{j})\label{eq:1}\end{eqnarray}

\item The pattern $\mathcal{E}=\left\{ E_{i}\right\} $ is inverse-stable
\emph{iff} there exists a partition $\mathcal{F}=\left\{ F_{j}^{+},F_{j}^{-}\right\} $
such that \begin{equation}
\bigoplus_{1\le i\le r}\mathbb{C}\widehat{\chi{(E}_{i})}=\bigoplus_{1\le j\le r}\mathbb{C}\left(\chi{(F}_{j}^{+})-\chi{(F}_{j}^{-})\right)\label{eq:2}\end{equation}

\item The signed-pattern $\mathcal{E}=\left\{ E_{i}^{+},E_{i}^{-}\right\} $
is product-stable \emph{iff} there exists a partition $\mathcal{F}=\left\{ F_{i}\right\} $
such that \begin{equation}
\bigoplus_{1\le i\le r}\mathbb{C}\left(\widehat{\chi(E_{i}^{+})}-\widehat{{\chi(E}_{i}^{-})}\right)=\bigoplus_{1\le j\le r}\mathbb{C}{\chi(F}_{j})\label{eq:3}\end{equation}

\item The signed-pattern $\mathcal{E}=\left\{ E_{i}^{+},E_{i}^{-}\right\} $
is inverse-stable \emph{iff} there exists a partition $\mathcal{F}=\left\{ F_{i}^{+},F_{i}^{-}\right\} $
such that \begin{equation}
\bigoplus_{1\le i\le r}\mathbb{C}\left(\widehat{\chi(E_{i}^{+})}-\widehat{{\chi(E}_{i}^{-})}\right)=\bigoplus_{1\le j\le r}\mathbb{C}\left(\chi{(F}_{j}^{+})-\chi{(F}_{j}^{-})\right)\label{eq:4}\end{equation}

\end{itemize}

\subsubsection{Stability by multiplication\label{sub:Stability-by-x_R2}}

If $\mathcal{E}$ and $\mathcal{F}$ are two signed-patterns verifying
Eq. \ref{eq:4} then for any $a$ prime with $q$ \[
a\mathcal{E}=\mathcal{E}\quad\mbox{and}\quad a\mathcal{F}=\mathcal{F}\]
 By $a\mathcal{E}=\mathcal{E}$ we mean $\forall i,\ \exists k$ such
that either $aE_{i}^{+}=E_{k}^{+}$and $aE_{i}^{-}=E_{k}^{-}$, or
$aE_{i}^{+}=E_{k}^{-}$ and $aE_{i}^{-}=E_{k}^{+}$.

\subsubsection{Admissible subsets\label{sub:Admissible-subsets_R3}}

All the sets $E_{i}^{\pm}$ or $E_{i}$ in relations Eq.\ref{eq:1}-Eq.\ref{eq:4}
are admissible.

\subsubsection{Convenient sets : necessary conditions of stability\label{sub:Convenient-sets_R4}}

This result is mainly useful for the demonstration of the result of
the paragraph \ref{sub:q-squareprime_R7}.

For a partition $\left\{ E_{1},\cdots,E_{r},A\right\} $ we define
a partition $\left\{ F_{1},\cdots,F_{s}\right\} $ such that $\bigoplus_{1\le i\le r}\mathbb{C}\widehat{\chi(E_{i})}\subset\bigoplus_{1\le j\le s}\mathbb{C}\chi(F_{j})$,
with $F_{j}$ maximal. We define $J$ the subset $\left\{ 1,\cdots,s\right\} $
by\[
j\in J\Longleftrightarrow\left(\widehat{\chi(E_{i})}\right)_{k}\ne0\quad\forall i\in\left\{ 1,\cdots,r\right\} ,k\in F_{j}\]
Then the set $\left\{ E_{1},\cdots,E_{r}\right\} $ is convenient
\emph{iff} \textcolor{black}{\[
\ \bigoplus_{j\in J}\mathbb{C}\widehat{\chi(F_{j})}\ \subset\left(\bigoplus_{1\le i\le r}\mathbb{C}\chi(E_{i})\right)\bigoplus\left(\bigoplus_{a\in A}\mathbb{C}\chi(\left\{ a\right\} )\right)\ \]
}

\subsubsection{Subgroups induce product-stable patterns\label{sub:subgroup_R5}}

If $H$ is a subgroup of $\mathbb{Z}_{q}^{\star}$ (the set of the
invertible elements of $\mathbb{Z}_{q}$) then $H$ induces a product-stable
pattern given by the classes modulo $H$ (i.e. the$\left\{ iH\right\} \ i\in\mathbb{Z}_{q}$)

\subsubsection{Monocolor inverse-stable signed-pattern\label{sub:Monocolor-inverse-stable-signed-patterns_R10}}

Except a $q=4$ (and $q=1$) example, there is \emph{no} inverse-stable
signed-pattern. This result proves a conjecture concerning cyclic
Hadamard matrices in the particular case of symmetric matrices.

\subsection{Proof and illustration of the above result\label{sub:Proof-or-justificationA}.}

Below we prove and comment the results mentioned above. We also give
examples and illustrations. First we need two assertions.

\begin{flushleft}\textbf{Assertion 1}: \emph{Let $V\subset\mathbb{C}^{q}$
be a vector subspace of dimension $r$ then $V$ is product-stable}
iff \emph{there exists disjoints subsets $F_{1},\cdots,F_{r}$ of
$\left\{ 1,\cdots,q\right\} $ with $V=\bigoplus_{1\le i\le r}\mathbb{C}\chi(F_{i})$. }\par\end{flushleft}

\begin{flushleft}\textbf{Assertion} \textbf{2}: \emph{Let $V\subset\mathbb{C}^{q}$
be a vector subspace of dimension $r$ and $V^{\star}$ the (supposed
nonempty) subset of vectors of $V$ with all non-zero components,
then $\left(V^{\star}\right)^{-1}\subset V$} iff \emph{there exist
a partition of $\left\{ 1,\cdots,q\right\} $ $F_{1}^{+},\ F_{1}^{-},\cdots,\ F_{r}^{+},\ F_{r}^{-}$
with} \[
V=\bigoplus_{1\le i\le r}\mathbb{C}\left(\chi(F_{i}^{+})-\chi(F_{i}^{-})\right)\]
\par\end{flushleft}

In other words Assertion 1 states that $V$ can be generated by vectors
with entries 0 or 1, and Assertion 2 states that $V$ can be generated
by vectors with entries 0, 1 or -1, and such that the absolute value
of all the entries of these vectors sum up to the all one entry vector.
These assertions are proved by recurrence on the space dimension $q$.

\paragraph*{Proof of Assertion 1}

For $q=1$ the assertion is clear. Suppose Assertion 1 is true for
dimension $q-1$ and let $V\in\mathbb{C}^{q}$ be a product-stable
subspace. Let $V\cap(\mathbb{C}^{q-1}\times\left\{ 0\right\} )=W\times\left\{ 0\right\} $
therefore $W=\oplus_{1\le l\le s}\mathbb{C}\chi(F_{l})$ where the
$F_{l}$ are disjoint subsets of $\left\{ 1,\cdots,q-1\right\} $.
Let $\mathbf{v}\in V$, we will show that $i,j\in F_{l}$ implies
$v_{i}=v_{j}$. If $v_{q}=0$ it is clear. If not, one can always
admit $v_{q}=1$, then $\mathbf{v}\cdot\mathbf{v}-\mathbf{v}\in W\times\left\{ 0\right\} $
so that $v_{i}^{2}-v_{i}=v_{j}^{2}-v_{j}$, which implies $v_{i}=v_{j}$
since $v_{i}+v_{j}=1$ is impossible (one can add $\chi(F_{l})\times\left\{ 0\right\} $
to $\mathbf{v}$). So if $\mathbf{v}\in V$ is a vector such that
$v_{q}=1$, we can admit $v_{i}=0$ for $i\in F_{l}\quad\forall l$.
Since $V=(W\times\left\{ 0\right\} )\oplus\mathbb{C}\mathbf{v}$ and
$\mathbf{v}\cdot\mathbf{v}=\mathbf{v}$ then all $\mathbf{v}=\chi(F)$
where $F\subset\left\{ 1,\cdots,q\right\} $ disjoint of all $F_{l}$\emph{.}

\paragraph*{Proof of Assertion 2}

Let $V$ satisfying the conditions of Assertion 2 and $\mathbf{v}\in V^{\star}$
such that $v_{q}=1$, we define $W=\left\{ \mathbf{w}\in V\mid w_{q}=0\right\} $
and $I=\left\{ i\mid w_{i}=0\;\forall\mathbf{w}\in W\right\} $. If
$I=\left\{ 1,\cdots,q\right\} $ then $W=\mathbb{C}\mathbf{v}$ therefore
$\mathbf{v}^{-1}=\mathbf{v}$ which proves the assertion. On the contrary
if $I\ne\left\{ 1,\cdots,q\right\} $ there exists $\mathbf{w}\in W$
such that $w_{i}\ne0$ for all $i\notin I$, this is possible while
$W$ is not a finite union of proper subspaces. Let $\mathbf{u}=\lim_{\epsilon\rightarrow0}\left(\mathbf{v}+\epsilon^{-1}\mathbf{w}\right)^{-1}$,
$u_{i}=v_{i}^{-1}$ if $i\in I$ and $u_{i}=0$ if $i\notin I$. On
another hand $\mathbf{v}-\mathbf{v}^{-1}\in W$ so $v_{i}=v_{i}^{-1}$
for $i\in I$, finally $v_{i}=\pm1$ for $i\in I$. One has $V=\mathbb{C}\mathbf{u}\oplus W$,
and we proceed again with $W$, keeping only the coordinates not in
$I$.

\subsubsection{Proof of \ref{sub:Patt-Subalg_R1}: pattern stability as matrix subalgebra}

Eq. \ref{eq:1} and Eq. \ref{eq:3} are direct consequences of the
assertion 1, taking \[
V=\bigoplus_{1\le i\le r}\mathbb{C}\widehat{\chi(E_{i})}\]
Eq. \ref{eq:2} and Eq. \ref{eq:4} are direct consequence  the assertion
2. If Eq. \ref{eq:1} holds, then $\left\{ 0\right\} \in\mathcal{E}$
and $\left\{ 0\right\} \in\mathcal{F}$. Indeed if $k\ne0$ and $0$
are in the same $F_{j}$ and $1\in E_{i}$ then $\left(\widehat{\chi(E_{i})}\right)_{0}=\left(\widehat{\chi(E_{i})}\right)_{k}$
is impossible.

We give in Appendix \ref{app:q8} the exhaustive list of the$P$-stable,
$I$-stable and $I\bar{P}$-stable pattern for $q=8$. The pattern
9 is an illustration of \ref{eq:1}, the pattern 5 is an illustration
of \ref{eq:3}, the pattern 18 is an illustration of \ref{eq:2},
and finally pattern 8 is an illustration of \ref{eq:4}. 

Note that inverse stability is the justification of introducing signed-pattern,
which could not be justified in the strict framework of lattice statistical
mechanics since asking that two Boltzmann weights are opposite is
unphysical (however such opposite entries in the Boltzman weight matrix
occurred in the solution of the $3d$ generalization of the Yang-Baxter
equation (tetrahedron equations) by Baxter and Zamolochikov).

In Appendix \ref{app:q8} we give a detailed non trivial example of
application of Eq. \ref{eq:4}.

\subsubsection{Proof of \ref{sub:Stability-by-x_R2}: stability by multiplication}

\textcolor{black}{Let $\mathcal{E}$ and $\mathcal{F}$ be two signed-patterns
verifying Eq. \ref{eq:4} and $A=\bigoplus_{1\le i\le r}\mathbb{C}\left(\widehat{\chi(E_{i}^{+})}-\widehat{{\chi(E}_{i}^{-})}\right)=\bigoplus_{1\le j\le r}\mathbb{C}\left(\chi{(F}_{j}^{+})-\chi{(F}_{j}^{-})\right)$
therefore for any $i$, $\widehat{\chi(E_{i}^{+})}-\widehat{{\chi(E}_{i}^{-})}\in A$.
This implies that, for any $j$, $\left(\widehat{\chi(E_{i}^{+})}\right)_{k}=\epsilon\left(\widehat{{\chi(E}_{i}^{-})}\right)_{l}$for
$k,l\in F_{j}^{+}\cup F_{j}^{-}$ with $\epsilon=1$ if $k$ and $l$
are both in $F_{j}^{+}$ or both in $F_{j}^{-}$, and $\epsilon=-1$
else. Let us introduce the polynomial\[
P(X)=\left(\sum_{e\in E_{i}^{+}}X^{ek}-\sum_{e\in E_{i}^{-}}X^{ek}\right)-\epsilon\left(\sum_{e\in E_{i}^{+}}X^{el}-\sum_{e\in E_{i}^{-}}X^{el}\right)\in\mathbb{Q\left[X\right]}\]
 One has} $P(\omega)=0$ and consequently, \textcolor{black}{using
a Galois symmetry argument,} $P(\omega^{a})=0$ for $a$ prime with
$q$. Therefore $\widehat{\chi(aE_{i}^{+})}-\widehat{{\chi(aE}_{i}^{-})}\in A$
and $\bigoplus_{1\le i\le r}\mathbb{C}\left(\widehat{\chi(E_{i}^{+})}-\widehat{{\chi(E}_{i}^{-})}\right)\subset A$.
The equality follows by a dimension argument. 

Notice that, applying this result to $a=-1$, one finds that if $E\in\mathcal{E}$
then either $E=-E$ or $-E$ is another set of $\mathcal{E}$.

\subsubsection{Proof of \ref{sub:Admissible-subsets_R3}: admissible subsets}

Let us first recall that an admissible set $E$ of $\left\{ 0,\cdots,q-1\right\} $
is a disjoint union

\[
E=\bigsqcup_{d\in D}d\ i_{d}H_{d}\]
where $D$ is a subset of the divisors of $q$, $H_{d}$ is a subgroup
of $\mathbb{Z}_{\frac{q}{d}}^{\star}$ and $\mbox{gcd}(di_{d},q)=d\;\forall d\in D$. 

We first note that the intersection of two admissible sets is an admissible
set. This comes from the fact that if $d$ is a divisor of $q$, $H$
and $H'$ are two subgroups of $\mathbb{Z}_{\frac{q}{d}}^{\star}$
, and $id$ and $i'd$ are two elements of $\mathbb{Z}(q,d)$ then
either there exists $i''d\in idH\cap i'dH'$ implying that $idH\cap i'dH'=i''d\left(H\cap H'\right)$,
or $idH\cap i'dH'=\emptyset$. We need also the following technical
lemma:

\begin{flushleft}\textbf{Lemma 1} L\emph{et $P(X)\in\mathbb{Z}\left[X\right]$,
if $id\in\mathbb{Z}(q,d)$ then $\left\{ jd\in\mathbb{Z}(q,d)\mid P(\omega^{id})=P(\omega^{jd})\right\} =idH$
with $H<\mathbb{Z}_{\frac{q}{d}}^{\star}$(with a little abuse of
notation).}\par\end{flushleft}

\begin{flushleft}\emph{Proof:} $\omega^{id}$ is a $\frac{q}{d}$th
primitive root of unity that we denote $\zeta$. Let $k$ be the inverse
of $i$ modulo $\frac{q}{d}$. The condition $P(\omega^{id})=P(\omega^{jd})$
becomes $P(\zeta)=P(\zeta^{kj})$. But the set $H=\left\{ m\in\mathbb{Z}_{\frac{q}{d}}^{\star}\mid P(\zeta)=P(\zeta^{m})\right\} $
is a subgroup of $\mathbb{Z}_{\frac{q}{d}}^{\star}$. Indeed $P(\zeta)=P(\zeta^{m})$
does not depend of the particular choice of the primitive root of
unity since it amounts to saying that $P(X)-P(X^{m})$ is a multiple
of the minimal polynomial of $\zeta$ and therefore $P(\zeta)=P(\zeta^{m})$
and $P(\zeta)=P(\zeta^{l})$ implies $P(\zeta)=P(\zeta^{ml})$ and
$P(\zeta)=P(\zeta^{m^{-1}})$.\par\end{flushleft}

Let us consider Eq. \ref{eq:4}. Let $k\in F_{j}^{+}$ and define
$P_{i}(X)=\sum_{e\in E_{i}^{+}}X^{e}-\sum_{e\in E_{i}^{-}}X^{e}$
. By Eq. \ref{eq:4} one has \[
F_{j}^{+}=\bigcap_{1\le i\le r}\left\{ l\in\mathbb{Z}_{q}\mid P_{i}(\omega^{l})=P_{i}(\omega^{k})\right\} \]
if $d=\gcd(q,k)$ then $F_{j}^{+}\cap\mathbb{Z}(q,d)=\bigcap_{1\le i\le r}\left\{ l\in\mathbb{Z}(q,d)\mid P_{i}(\omega^{l})=P_{i}(\omega^{k})\right\} $
which is an admissible set by the results above, and finally $F_{j}^{+}$
is also admissible.

\subsubsection{Proof of \ref{sub:Convenient-sets_R4}: convenient sets, necessary
conditions of stability}

We define the intersection of two partitions $\mathcal{E}=\left\{ E_{1},\cdots,E_{r}\ \right\} $
and $\mathcal{F}=\left\{ F_{1},\cdots,F_{s}\right\} $ of a finite
set $I$ by\[
\mathcal{E}\cap\mathcal{F}=\left\{ E_{i}\cap F_{j}\left|1\le i\le r\right.\ 1\le j\le s\right\} \]

\begin{flushleft}\textbf{Lemma} 2 \emph{Let $\mathbf{y}\in\mathbb{C}^{n}$and
$\mathbf{a(1)},\cdots,\mathbf{a(t)}\in\mathbb{C}^{n}$ for $A\in\mathcal{P}_{\mathbf{a}(1)}{\cap\cdots\cap\mathcal{P}}_{\mathbf{a}(t)}$.
The two following affirmations \ref{eq:unconv1} and \ref{eq:unconv2}
are equivalent (with the convention $0^{0}=1$)}\begin{equation}
\forall(k_{1},\cdots,k_{t})\in\mathbb{N}^{t}\qquad\sum_{i=1}^{n}a(1)_{i}^{k_{1}}\cdots a(t)_{i}^{k_{t}}y_{i}=0\label{eq:unconv1}\end{equation}
\textcolor{black}{\begin{equation}
\forall A\in\mathcal{P}_{\mathbf{a(1)}}\cap\cdots\cap\mathcal{P}_{\mathbf{a(t)}}\qquad\sum_{i\in A}y_{i}=0\label{eq:unconv2}\end{equation}
}\par\end{flushleft}

\begin{flushleft}\emph{Proof: } The proof goes by induction over $t.$
For $t=1$, let us define the sets $A_{i}$ by $\mathcal{P}_{\mathbf{a(1)},\left\{ 1,\cdots,n\right\} }=\left\{ A_{1},\cdots,A_{r}\right\} $
and let  $b_{j}$ be the value of $\mathbf{a(1)}_{i}$ for $i$ in
$A_{j}$. Using \ref{eq:unconv1} one has\begin{equation}
\begin{array}{cc}
\underbrace{\left(\begin{array}{ccc}
1 & \cdots & 1\\
\cdots & \cdots & \cdots\\
b_{1}^{r-1} & \cdots & b_{r}^{r-1}\end{array}\right)} & \left(\begin{array}{c}
\sum_{i\in A_{1}}y_{i}\\
\vdots\\
\sum_{i\in A_{r}}y_{i}\end{array}\right)=\left(\begin{array}{c}
0\\
\vdots\\
0\end{array}\right)\\
B\end{array}\label{eq:cadix}\end{equation}
Since $\det B=\prod_{1\le i<j\le r}(b_{i}-b_{j})\ne0$ (Vandermonde
determinant), $\sum_{i\in A_{k}}y_{i}=0$ for any $1\le k\le r$.
This proves the property for $t=1$.\par\end{flushleft}

Let us take $t>1$ and assume the lemma for $t-1$, therefore Eq.
\ref{eq:unconv1} is equivalent to\[
\sum_{i\in A}{a(t)}_{i}^{k_{t}}y_{i}=0\quad\forall A\in\mathcal{P}_{\mathbf{a(1)}}\cap\cdots\cap\mathcal{P}_{\mathbf{a(t-1)}}\quad\forall k_{t}\in\mathbb{N}\]
and using the lemma for $t=1$ this is equivalent to\[
\sum_{i\in C}y_{i}=0\quad{\forall C\in\mathcal{P}}_{\mathbf{a(t)},A}\ \mbox{with }A\in\mathcal{P}_{\mathbf{a(1)}}\cap\cdots\cap\mathcal{P}_{\mathbf{a(t-1)}}\]
\textcolor{black}{\[
\sum_{i\in C}y_{i}=0\quad C\in\mathcal{P}_{\mathbf{a(1)}}\cap\cdots\cap\mathcal{P}_{\mathbf{a(t)}}\]
}which completes the proof of the lemma.

We now use this lemma to prove the result \ref{sub:Convenient-sets_R4}.
Let $\left\{ E_{0},E_{1},\cdots,E_{r},A\right\} $ be a partition
such that $\mathcal{E}=\left\{ E_{0},E_{1},\cdots,E_{r}\right\} $
is convenient. For $(k_{0},\cdots,k_{r})\in\mathbb{N}^{r+1}$ one
introduces $\mathbf{u}=\chi(E_{0})^{\star k_{0}}\cdots\chi(E_{r})^{\star k_{r}}$.
By definition $u_{i}=u_{j}$ for $i$ and $j$ in the same $E_{l}$.
If $\mathbf{v}=\widehat{\mathbf{u}}=q^{k_{0}+\cdots+k_{r}}\widehat{\chi(E_{0})}^{k_{0}}\cdots\widehat{\chi(E_{r})}^{k_{r}}$this
implies $\sum_{m}v_{m}\omega^{-im}=\sum_{m}v_{m}\omega^{-jm}$ leading
to\[
\sum_{m}\left(\omega^{-im}-\omega^{-jm}\right)\left(\widehat{\chi(E_{0})}\right)_{m}^{k_{0}}\cdots\left(\widehat{\chi(E_{r})}\right)_{m}^{k_{r}}=0\]
We now use the lemma 2 with $y_{m}=\omega^{-im}-\omega^{-jm}$ and
$\mathbf{a}(n)=\widehat{\chi(E_{n})}$ to get the result.

\subsubsection{Proof of \ref{sub:subgroup_R5}: subgroups induce product-stable
patterns}

To show that subgroups induce product-stable pattern, we first note
that the class modulo $H$ is indeed a pattern. Let $I$ be a set
of representative of the classes modulo $H$. Let \[
\mathbf{x}=\sum_{i\in I}a_{i}\chi(iH)\in\bigoplus_{i\in I}\mathbb{C}\chi(iH)\]
then\textcolor{black}{\[
\left(\widehat{\mathbf{x}}\right)_{j}=\sum_{i\in I}a_{i}\frac{\left|iH\right|}{\left|H\right|}\sum_{h\in H}\omega^{ihj}\]
where $\left|A\right|$ denotes the cardinality of the set $A$. C}onsequently
if $j'=tj$ with $t\in H$ then $\left(\widehat{\mathbf{x}}\right)_{j}{=\left(\widehat{\mathbf{x}}\right)}_{j'}$
which implies\[
\bigoplus_{i\in I}\mathbb{C}\widehat{\chi(iH)}\subset\bigoplus_{i\in I}\mathbb{C}\chi(iH)\]
The inverse inclusion is shown using inverse Fourier transform.

\subsubsection{Proof of \ref{sub:Monocolor-inverse-stable-signed-patterns_R10}:
monocolor inverse-stable signed-pattern}

Let us take $q>1$ (the case $q=1$ is obvious). It is readily verified
that the monocolor signed-pattern $\left[a,-a,-a,-a\right]$ is $I$-stable.
We now prove the following lemma which we will need.

\begin{flushleft}\textbf{Lemma 3} \emph{if $a,b\in\mathbb{N}^{\star}\ a\mathbb{Z}_{ab}^{\star}=a\mathbb{Z}_{b}^{\star}$(by
$\mathbb{Z}_{b}^{\star}$ we means the elements of $\mathbb{Z}_{ab}$
which are prime with $b$).} \par\end{flushleft}

\begin{flushleft}\emph{Proof: } Let us introduce the set $C=\left\{ k\in\mathbb{Z}_{ab}\mid\gcd(k,b)=1\right\} $,
we will prove $a\mathbb{Z}_{ab}^{\star}=aC$. It is clear that $a\mathbb{Z}_{ab}^{\star}\subset aC$.
We need to show that if $k$ is prime with $b$, then one of the $k,\ k+b,\cdots,k+(a-1)b$
is prime with $a$. Let us write $a=cd$ where the prime factors of
$b$ appear only in $c$. Since $\gcd(b,d)=1$ then $k,\ k+b,\cdots,\ k+(d-1)b$
are distinct modulo $d$, therefore one of them is equal to 1 modulo
$d$, which proves the lemma.\par\end{flushleft}

Below we show that there is no other $I$-stable signed-pattern than
$\left[a,-a,-a,-a\right]$. Let $E^{+}$, $E^{-}$ be an $I$-stable
monocolor signed-pattern and $M=\mbox{Cy}\left(\chi(E^{+})-\chi(E^{-})\right)$,
the inverse-stability can be expressed as $M^{2}=tI_{q}$ where $t$
is some even non zero integer and $I_{q}$ is the identity matrix.
Applying twice $M$ to the all-one entry vector, one gets\begin{equation}
M^{2}=s^{2}I_{q}\label{eq:M2}\end{equation}
with $s=\left|E^{+}\right|-\left|E^{-}\right|$ where we consider,
without loss of generality, that $\left|E^{+}\right|>\left|E^{-}\right|$. 

We now prove that $s^{2}=q$. Indeed using Eq. \ref{eq:4} \[
U^{\star}MU=q\mbox{ diag}\left(\widehat{\chi(E^{+})}-\widehat{\chi(E^{-})}\right)\]
so there exists a constant $k$ and a partition $\left\{ F^{+},F^{-}\right\} $
of $\left\{ 0,\cdots,q-1\right\} $ such that $\widehat{\chi(E^{+})}-\widehat{\chi(E^{-})}=k\left(\chi(F^{+})-\chi(F^{-})\right)$.
$s=\left(\widehat{\chi(E^{+})}-\widehat{\chi(E^{-})}\right)_{0}=k\left(\chi(F^{+})-\chi(F^{-})\right)_{0}$
so $s=k$\begin{equation}
\widehat{\chi(E^{+})}-\widehat{\chi(E^{-})}=s\left(\chi(F^{+})-\chi(F^{-})\right)\label{eq:monoc}\end{equation}
Define $N=\mbox{Cy}\left(\chi(F^{+})-\chi(F^{-})\right)$ applying
again the Fourier transform to Eq. \ref{eq:monoc} one gets \[
N^{2}=\left(\frac{q}{s}\right)^{2}I_{q}\]
which combined with Eq. \ref{eq:M2} yields \[
M^{2}=qI_{q}\]
as stated before.

Applying the equation above to a diagonal term proves that $M$ is
symmetric and therefore Eq. \ref{eq:M2} can be written as \begin{equation}
\widetilde{M}M=qI_{q}\label{eq:triv}\end{equation}
In other words, $M$ is a so-called \cite{CyclotomicInteger} symmetric
Hadamard matrix, see ref \cite{Hadamardmatrices}. 

Using Eq. \ref{eq:M2} and Eq. \ref{eq:triv} one gets $q=4u^{2}$.
The example given in the beginning of this paragraph corresponds to
$u=1$. 

Since $\widehat{\chi(E^{+})}-\widehat{\chi(E^{-})}=2\widehat{\chi(E^{+})}-\widehat{\chi(\left\{ 0,\cdots,q-1)\right\} )}=2\widehat{\chi(E^{+})}-q\chi(\left\{ 0)\right\} )$
and using Eq. \ref{eq:monoc} \begin{equation}
\left(\widehat{\chi(E^{+})}\right)_{i}=\pm u\quad\mbox{for }i\ne0\label{eq:pm}\end{equation}
 Since $E^{+}$ is an admissible set (see paragraph \ref{sub:Admissible-subsets_R3})
\[
E^{+}=\bigsqcup_{d\in D}di_{d}H_{d}\]
where $D$ is a subset of the divisors of $q$, ${\mbox{gcd}(d\ i}_{d},q)=d$
and $H_{d}<\mathbb{Z}_{\frac{q}{d}}^{\star}$. Using the result of
subsection \ref{sub:Stability-by-x_R2} and since $\left|E^{+}\right|\ne\left|E^{-}\right|$
for any $a\in\mathbb{Z}_{q}^{\star}$ one has $aE^{+}=E^{+}$ yielding
$\mathbb{Z}_{q}^{\star}E^{+}=E^{+}$, in particular for $d\in D$,
$i_{d}d\mathbb{Z}_{q}^{\star}\subset E^{+}$. Using Lemma 3 one has
$i_{d}d\mathbb{Z}_{q}^{\star}=i_{d}d\mathbb{Z}_{\frac{q}{d}}^{\star}=d\mathbb{Z}_{\frac{q}{d}}^{\star}$,
so that\begin{equation}
E^{+}=\bigsqcup_{d\in D}d\mathbb{Z}_{\frac{q}{d}}^{\star}\label{eq:reu}\end{equation}
(in the example shown in the beginning of this section one has $E^{+}=\left\{ 1,3\right\} \sqcup\left\{ 2\right\} $)
.

At this point we need to use results of number theory. The so-called
Moebius function $\mu$ \cite{moebiusfunction} is defined by \[
\mu(n)=\left\{ \begin{array}{cc}
1 & \quad\mbox{if }n=1\\
(-1)^{l} & \quad\mbox{if }n=p_{1}\cdots p_{l}\\
0 & \quad\mbox{else}\end{array}\quad\mbox{with }p_{1},\cdots,p_{l}\mbox{ distinct primes}\right.\]
and it has the property that $\sum_{k\in\mathbb{Z}_{n}^{\star}}\zeta^{k}=\mu(n)$
where $\zeta=\exp\frac{2\pi}{n}\imath$. In our case one gets $\sum_{k\in\mathbb{Z}_{q/d}^{\star}}\omega^{kd}=\mu(\frac{q}{d})$.
We now use Eq. \ref{eq:pm} and Eq. \ref{eq:reu} and we get \[
\pm u=\left(\widehat{\chi(E^{+})}\right)_{1}=\sum_{d\in D}\mu(\frac{q}{d})\]
Noting $q=2^{2a_{0}}p_{1}^{2a_{1}}\cdots p_{l}^{2a_{l}}$where $2,{\ p}_{1},\cdots,{\ p}_{l}$
are distinct prime numbers, one has\[
\left|\left\{ t|q\;\mbox{such that }\mu(t)=1\right\} \right|=2^{l}=\left|\left\{ t|q\;\mbox{such that }\mu(t)=-1\right\} \right|\]
therefore $u=2^{a_{0}-1}p_{1}^{a_{1}}\cdots p_{l}^{a_{l}}\le2^{l}$,
consequently $l=0$ and $a_{0}=1$, which proves that $q=4$ is the
only possible value. 

%
\begin{comment}
There is a conjecture ref stating that there there is no cyclic Hadamard
matrices for $q>4$. This has been proved for $q<4\ 10^{11}$, with
possible exceptions of $q=108900,\ 47436840000,\ 26899280100$.
\end{comment}
{}

\section{Analytical results pertaining to statistical mechanics \label{sec:Analytical-Results-Meca-stat}}

In this paragraph we express the inverse-stability and the product-stability
for the pattern and the signed-pattern in term of matrix subalgebra
and list our analytical results.

\subsection{List of the results\label{sub:List-of-theB}}

\subsubsection{$q$ is a prime\label{sub:q-prime_R6}}

When $q$ is a prime number, there is no other product-stable pattern
than the patterns induced by the subgroups of $\mathbb{Z}_{q}$. Furthermore
there is no inverse-stable pattern which is not product-stable. Consequently
there are $1+\tau(q-1)$ stable patterns, where $\tau(n)$ is the
number of divisors of $n$.

\subsubsection{$q$ is the square of a prime or the product of two primes\label{sub:q-squareprime_R7}}

If $q$ is the square of a prime $p$, then there are $1+\tau(p-1)+\tau^{2}(p-1)$
product-stable patterns. In addition to the class modulo $Z_{q}$
described in \ref{sub:subgroup_R5}, there exists other product-stable
patterns. If $P$ denotes the natural projection from $\mathbb{Z}_{q}$
on $\mathbb{Z}_{p}$ the set of product-stable patterns is: \[
\left\{ \left\{ 0\right\} ,\mathbb{Z}_{q}\setminus\left\{ 0\right\} \right\} \bigcup\left\{ \alpha L\right\} \bigcup\left\{ \left\{ 0\right\} ,aP^{-1}(H),bpK\right\} \]
where\begin{eqnarray*}
\alpha\in\mathbb{Z}_{q} &  & L<\mathbb{Z}_{q}\\
a,b\in\mathbb{Z}_{p}^{\star} &  & H,K<\mathbb{Z}_{p}^{\star}\end{eqnarray*}

If $q=p_{1}p_{2}$ is the product of two different prime numbers $p_{1}$
and $p_{2}$ then the three-color pattern defined by $E_{0}=\left\{ 0\right\} $
and $E_{1}=\left\{ p_{1},2p_{1},\cdots,(p_{2}-1)p_{1}\right\} $ is
product-stable.

\subsubsection{An integrable one-parameter family of integrable patterns\label{sub:inte-fam_R8}}

if $q\ne2$ is a prime the three-color pattern formed by\begin{eqnarray}
E_{0} & = & \left\{ 0\right\} \nonumber \\
E_{1} & = & \left\{ i^{2},i\in\mathbb{Z}_{q}^{\star}\right\} \label{eq:int}\\
E_{2} & = & \mathbb{Z}_{q}-E_{0}-E_{1}\nonumber \end{eqnarray}
is product-stable. Furthermore if $q=4k+1$, then the associated transformation
$K$ is integrable.

\subsubsection{Six families of stable patterns\label{sub:3coul-genera_R9}}

We give below six families of patterns which are stable for any even
$q$. Two patterns on the same row of the table below are related
by discrete Fourier transform. \begin{eqnarray*}
\mbox{Prod.-stable signed-pattern} &  & \mbox{Inv.-stable simple-pattern}\\
P_{1}=\left[x_{0},x_{1},\cdots,-x_{1},\cdots,x_{1}\right] &  & Q_{1}=\left[x_{0},\cdots,x_{0},x_{1},x_{0},\cdots,x_{0}\right]\\
P_{2}=\left[x_{0},x_{1},-x_{1},\cdots,x_{2},-x_{1},\cdots,x_{1}\right] &  & Q_{2}=\left[x_{0},x_{1},x_{0},\cdots,x_{0},x_{2},x_{0},x_{1},\cdots,x_{1}\right]\\
P_{3}=\left[x_{0},x_{1},x_{2},\cdots,x_{\frac{q}{2}},-x_{1},\cdots,-x_{\frac{q}{2}-1}\right] &  & Q_{3}=\left[x_{0},x_{1},x_{0},x_{2},x_{0},x_{3},\cdots,x_{0},x_{\frac{q}{2}}\right]\end{eqnarray*}
Patterns $P_{1}$ and $Q_{1}$ are two-color pattern, $P_{2}$ and
$Q_{2}$ are three-color pattern , and $P_{3}$ and $Q_{3}$ are $\frac{q}{2}+1$-color
patterns. For pattern $P_{1}$ (resp. $Q_{1}$) the entry$-x_{1}$
(resp. $x_{1}$) is in position $\frac{q}{2}+1$ (position starting
at zero). For pattern $P_{2}$ and $Q_{2}$ the entry $x_{2}$ is
also in position $\frac{q}{2}+1$. Note that for $P_{2}$ the elements
before and after $x_{2}$ are $x_{1}$when $\frac{q}{2}$ is even
and $-x_{1}$ when $\frac{q}{2}$ is odd.

\subsection{Proof and illustration of the above results\label{sub:Proof-or-justificationB}.}

\subsubsection{Proof of \ref{sub:q-prime_R6}: q is a prime}

The key point of this demonstration is the well known fact that (if
$a_{0},\cdots,a_{q-1}\in\mathbb{Q}$) \begin{equation}
\sum_{i=0}^{q-1}a_{i}\omega^{i}=0\Longleftrightarrow a_{0}=\cdots=a_{q-1}\label{eq:polyannul}\end{equation}
Let $\mathcal{E}=\left\{ E_{1}^{+},E_{1}^{-},\cdots,E_{r}^{+},E_{r}^{-}\right\} $
and $\mathcal{F}=\left\{ F_{1}^{+},F_{1}^{-},\cdots,F_{r}^{+},F_{r}^{-}\right\} $
be two partitions verifying Eq. \ref{eq:4} with $E_{1}^{+},\cdots,E_{r}^{+}$,
$F_{1}^{+},\cdots,F_{r}^{+}$non empty. We admit $r\ge2$ since the
case $r=1$ occurs only for $q=1$ or $q=4$ as shown in section \ref{sub:Monocolor-inverse-stable-signed-patterns_R10}.
Note that the $E$'s and the $F$'s play the same role and can be
interchanged.

Let $1\le l,m\le r$, Eq. \ref{eq:4} can be rewritten as \begin{equation}
\left.\begin{array}{cc}
\left(\widehat{\chi(E_{l}^{+})}-\widehat{{\chi(E}_{l}^{-})}\right)_{i}=\left(\widehat{\chi(E_{l}^{+})}-\widehat{{\chi(E}_{l}^{-})}\right)_{j} & \mbox{if }i,j\in F_{m}^{+}\mbox{ or }i,j\in F_{m}^{-}\\
\left(\widehat{\chi(E_{l}^{+})}-\widehat{{\chi(E}_{l}^{-})}\right)_{i}=-\left(\widehat{\chi(E_{l}^{+})}-\widehat{{\chi(E}_{l}^{-})}\right)_{j} & \mbox{ if }i\in F_{m}^{+}\mbox{ and }j\in F_{m}^{-}\end{array}\right\} \label{eq:pe}\end{equation}

If $0\in F_{1}^{+}$we will show $F_{1}^{+}=\left\{ 0\right\} $ and
$F_{1}^{-}=\emptyset$. Suppose, \emph{ab absurdo}, that there exists
$i\ne0$ with $i\in F_{1}^{+}\cup F_{1}^{-}$. Using Eq. \ref{eq:pe}
one gets\[
\left(\widehat{\chi(E_{l}^{+})}-\widehat{{\chi(E}_{l}^{-})}\right)_{0}=\pm\left(\widehat{\chi(E_{l}^{+})}-\widehat{{\chi(E}_{l}^{-})}\right)_{i}\]
for $l\ne1$ and $0\in E_{1}^{+}$. So that \[
\left|E_{l}^{+}\right|-\left|E_{l}^{-}\right|=\pm\left(\sum_{e\in E_{l}^{+}}\omega^{ie}-\sum_{e\in E_{l}^{-}}\omega^{ie}\right)\]
which is impossible by Eq. \ref{eq:polyannul}, since $0\notin iE_{l}^{+}$
and $iE_{l}^{+}\cap iE_{l}^{-}=\emptyset$. The same applies interchanging
the $E$'s and $F$'s.

We now show that $E_{1}^{-}=\cdots=E_{r}^{-}=F_{1}^{-}=\cdots=F_{r}^{-}=\emptyset$.
Recalling that $E_{1}^{+}=\left\{ 0\right\} $ and $E_{1}^{-}=\emptyset$,
let suppose that $F_{l}^{-}$is non empty and $i\in F_{l}^{+}$, $j\in F_{l}^{-}$,
using Eq. \ref{eq:pe} one has $\left(\widehat{\chi(E_{1}^{+})}\right)_{i}={-\left(\widehat{\chi(E_{1}^{+})}\right)}_{j}$
and consequently $1=-1$, a contradiction.

So, when $q$ prime, Eq. \ref{eq:4} reduces to Eq. \ref{eq:1} which
we write \begin{equation}
\left(\widehat{\chi(E_{k}^{+})}\right)_{i}=\left(\widehat{\chi(E_{k}^{+})}\right)_{j}\quad\forall i,j\in F_{l},\ \forall k,l\label{eq:mp}\end{equation}
if $1\in H\in\mathcal{E}$ and $1\in K\in\mathcal{F}$ so $\sum_{h\in H}\omega^{ih}=\sum_{h\in H}\omega^{jh}$
for $i,j\in K$ . Using Eq. \ref{eq:polyannul}, one deduces $iH=jH$.
Taking $j=1$ one gets $i\in H$ and therefore $K\subset H$. Interchanging
$H$ and $K$ one finds that $K=H$. From $ij^{-1}\in H$ we deduce
that the subset of the partition which contains 1 is a subgroup.

Again using Eq. \ref{eq:mp} with $l\ge2$ one has $iE_{l}^{+}=jE_{l}^{+}$
for $i,j\in H$ and therefore $E_{l}^{+}=HE_{l}^{+}$and $E_{l}^{+}$
is an union of classes modulo $H$. On another hand again using Eq.
\ref{eq:mp}\[
\left(\widehat{\chi(H)}\right)_{i}=\left(\widehat{\chi(H)}\right)_{j}\quad\forall i,j\in E_{l}^{+}\]
therefore $iH=jH$ and $E_{l}^{+}$ is one class modulo $H$. This
completes the proof (noting $E_{1}^{+}=\left\{ 0\right\} =0H$).

\subsubsection{Proof of \ref{sub:q-squareprime_R7}: $q$ is the square of a prime}

We consider the case $q=p^{2}$ with $p$ a prime and note $\omega=\exp\frac{2\pi}{q}\imath$
and $\zeta=\exp\frac{2\pi}{p}\imath$. We recall that the cyclotomic
polynomial of respective order $p$ and $q$ are\begin{eqnarray*}
\Phi_{p}(X) & = & 1+\cdots+X^{p-1}\\
\Phi_{q}(X) & = & 1+X^{p}+\cdots+X^{(p-1)p}\end{eqnarray*}
therefore if $P(X)=\sum_{k=0}^{p-1}a_{k}X^{k}$ and $Q(X)=\sum_{k=0}^{p^{2}-1}b_{k}X^{k}$
are two polynomials in $\mathbb{Q}[X]$ then\begin{eqnarray}
P(\zeta)=0 & \Longleftrightarrow & a_{0}=\cdots=a_{p-1}\label{eq:cyclozeta}\\
Q(\omega)=0 & \Longleftrightarrow & \left(p\mid(l-m)\Rightarrow b_{l}=b_{m}\right)\label{eq:cyclomega}\end{eqnarray}
A minimal polynomial of $\omega$ over $\mathbb{Q}[\zeta]$ is $X^{p}-\zeta$.
Finally we note $P$ the natural projection $\mathbb{Z}_{q}\overset{P}{\rightarrow}\mathbb{Z}_{p}$.

\begin{flushleft}\textbf{Lemma 4} \emph{If $H<\mathbb{Z}_{q}^{\star}$
then either $H=P^{-1}(K)$ with $K<\mathbb{Z}_{p}^{\star}$ (in that
case $H$ is said to be of type 1) or $h\in H,\ h'\in H,\ h\ne h'$
implies $p\nmid(h'-h)$ (in that case $H$ is said to be of type 2).}\par\end{flushleft}

\begin{flushleft}\emph{Proof}: Suppose $H<\mathbb{Z}_{q}^{\star}$
is not of type 2, then there exists $h$, $h',$ $h'\ne h$ such that
$p\mid(h-h')$, so $h'h^{-1}=1+kp$ with $1\le k\le p-1$. Therefore
$(1+kp)^{l}=1+klp$ for all $l$ and $1+p\mathbb{Z}_{p}\subset H$
and finally $H(1+p\mathbb{Z}_{p})=H$: $H$ is of type 1.\par\end{flushleft}

The next lemma is devoted to the determination of $\mathcal{P}_{\widehat{\chi(E)}}$
when $E=iH$ ($p\nmid i$ and $H<\mathbb{Z}_{q}^{\star}$) or $E=jpK$
($p\nmid j$ and $K<\mathbb{Z}_{p}^{\star}$).

\begin{flushleft}\textbf{Lemma 5} \emph{One distinguishes the three
following cases}\begin{equation}
E=jpK,\ p\nmid j,\ K<\mathbb{Z}_{p}^{\star}\Longrightarrow\mathcal{P}_{\widehat{\chi(E)}}=\left\{ p\mathbb{Z}_{p},\left\{ iP^{-1}(K)\vert\ p\nmid i\right\} \right\} \label{eq:lm4_1}\end{equation}
\begin{equation}
E=iP^{-1}(K),\ K<\mathbb{Z}_{p}^{\star},\ p\nmid it\Longrightarrow\mathcal{P}_{\widehat{\chi(E)}}=\left\{ \left\{ 0\right\} ,\mathbb{Z}_{q}^{\star},\left\{ pjK\vert\ p\nmid j\right\} \right\} \label{eq:lm4_2}\end{equation}
\begin{equation}
E=iH,\ p\nmid i,\ H<\mathbb{Z}_{q}^{\star},\ H\mbox{ of type 2}\Longrightarrow\mathcal{P}_{\widehat{\chi(E)}}=\left\{ jH\vert\ j\in\mathbb{Z}_{q}\right\} \label{eq:lm4_3}\end{equation}
\par\end{flushleft}

\begin{flushleft}\emph{Proof of} \emph{Eq.} \ref{eq:lm4_1}\par\end{flushleft}

\begin{flushleft}If $p\nmid i$ and $p\nmid l$ then \[
\left(\widehat{\chi(E)}\right)_{i}=\left(\widehat{\chi(E)}\right)_{l}\Leftrightarrow\sum_{k\in K}\left(\omega^{ijpk}-\omega^{ljpk}\right)=0\Leftrightarrow\sum_{k\in K}\left(\zeta^{ijk}-\zeta^{ljk}\right)=0\overset{(1)}{\Longleftrightarrow}iK=lK\]
on another hand $\left(\widehat{\chi(E)}\right)_{pt}=\left|K\right|$
for any $t$. The equivalence noted $\overset{(1)}{\Longleftrightarrow}$
refers to \ref{eq:cyclozeta} with $H<\mathbb{Z}_{q}^{\star}\ p\nmid i$ \par\end{flushleft}

\begin{flushleft}\emph{Proof of Eq.} \ref{eq:lm4_2}\par\end{flushleft}

\begin{flushleft}If $K=\mathbb{Z}_{p}^{\star}$, then $E=\mathbb{Z}_{q}^{\star}$
and \ref{eq:lm4_2} is verified. We now suppose $K\ne\mathbb{Z}_{p}^{\star}$.
If $p\nmid l$ then $\left(\widehat{\chi(E)}\right)_{l}=\sum_{k\in K}\sum_{m=0}^{p-1}\omega^{il(mp+k)}=0$,
on another hand if $p\nmid t$ $\left(\widehat{\chi(E)}\right)_{pt}=p\sum_{k\in K}\zeta^{itk}\ne0$
using \ref{eq:cyclozeta} and $K\ne\mathbb{Z}_{p}^{\star}$, the value
$\sum_{k\in K}\zeta^{itk}$ depends only on $K$ using again \ref{eq:cyclozeta}.
Finally $\left(\widehat{\chi(E)}\right)_{0}=p\left|K\right|$.\par\end{flushleft}

\begin{flushleft}\emph{Proof of Eq.} \ref{eq:lm4_3} \par\end{flushleft}

\begin{flushleft}If $p\nmid l$ and $p\nmid j$ the two quantities
$\left(\widehat{\chi(E)}\right)_{j}=\sum_{h\in H}\omega^{ijh}$ and
$\left(\widehat{\chi(E)}\right)_{l}=\sum_{h\in H}\omega^{ilh}$ are
equal \emph{iff} $ijH=ilH$. Indeed let us suppose $\left(\widehat{\chi(E)}\right)_{j}=\left(\widehat{\chi(E)}\right)_{l}$
then $\forall k\in\left[1,p-1\right],\left|ijH\cap(k+p\mathbb{Z}_{p})\right|=0,\mbox{ or }1$
since $H$ is of type 2, and if $t\in\left(ijH\cap(k+p\mathbb{Z}_{p})\right)\setminus\left(ilH\cap(k+p\mathbb{Z}_{p})\right)$
then, using \ref{eq:cyclomega}, $t+p,\cdots,t+(p-1)p$ also belong
to $\left(ijH\cap(k+p\mathbb{Z}_{p})\right)\setminus\left(ilH\cap(k+p\mathbb{Z}_{p})\right)$,
a contradiction. On another hand $\left(\widehat{\chi(E)}\right)_{j}$
can be seen as a polynomial in $\omega$ of degree strictly smaller
than $p$ over $\mathbb{Q}\left[\zeta\right]$ therefore $\left(\widehat{\chi(E)}\right)_{j}\notin\mathbb{Q}\left[\zeta\right]$
(see above point 4 of the recalls). Finally if $p\nmid l$ and $p\nmid j$
the two quantities$\left(\widehat{\chi(E)}\right)_{pj}=\sum_{h\in H}\zeta^{ijh}\in\mathbb{Q}\left[\zeta\right]$
and $\left(\widehat{\chi(E)}\right)_{pl}=\sum_{h\in H}\zeta^{ilh}\in\mathbb{Q}\left[\zeta\right]$
are equal \emph{iff} $ijP(H)=ilP(H)$ using \ref{eq:cyclozeta}.\par\end{flushleft}

\begin{flushleft}\textbf{Lemma 6} \emph{If $E$ is convenient, admissible,
$E\cap\mathbb{Z}_{q}^{\star}\neq\emptyset$ and $E\cap p\mathbb{Z}_{p}^{\star}\neq\emptyset$
then $E=\mathbb{Z}_{q}\setminus\left\{ 0\right\} $.}\par\end{flushleft}

\begin{flushleft}\emph{Proof:} \textbf{} Let us take a set $E$ verifying
the conditions of the lemma, since $E$ is admissible then $E=iH\sqcup jpK$
with $H<\mathbb{Z}_{q}^{\star}$ , $K<\mathbb{Z}_{p}^{\star}$and
$p\nmid i$ and $p\nmid j$ . If there is $A\in\mathcal{P}_{\widehat{\chi(E)}}$
with $A=pM\subset p\mathbb{Z}_{p}^{\star}$ using ({*}) $\left(\widehat{\chi(A)}\right)_{i}=\left(\widehat{\chi(A)}\right)_{jp}$
, therefore $\sum_{m\in M}\omega^{imp}=\sum_{m\in M}\omega^{jmp^{2}}$,
so $\sum_{m\in M}\zeta^{im}=\left|M\right|$ which is impossible.
Consequently there exists $t$ relatively prime to $p$ such that
$t,p\in B\in\mathcal{P}_{\widehat{\chi(E)}}$ and so $\left(\widehat{\chi(E)}\right)_{t}=\left(\widehat{\chi(E)}\right)_{p}$
which reads \begin{equation}
\sum_{h\in H}\omega^{iht}+\sum_{k\in K}\zeta^{jkt}=\sum_{h\in H}\zeta^{ih}+\sum_{k\in K}1\label{eq:comme}\end{equation}
So $\sum_{h\in H}\omega^{iht}\in\mathbb{Q}\left[\zeta\right]$ and
$H$ cannot be of type 2 since the degree of $\omega$ over $\mathbb{Q}\left[\zeta\right]$
is $p$. $H$ is then of type 1, $H=P^{-1}(L)$ where $L<\mathbb{Z}_{p}^{\star}$.
Using \ref{eq:comme}, $\sum_{k\in K}\zeta^{jkt}=p\sum_{l\in L}\zeta^{il}+\left|K\right|$
. We use \ref{eq:cyclozeta} and we note $A_{s}=\left|\left\{ k\in K\vert p\mid jkt-s\right\} \right|$
and $B_{s}=\left|\left\{ l\in L\vert p\mid jl-s\right\} \right|$
for $1\le s<p$, one deduces $\left|K\right|=-A_{s}+pB_{s}$ which
is possible only if $A_{s}=B_{s}=1\quad\forall s$, therefore $K=L=\mathbb{Z}_{p}^{\star}$
and finally $E=\mathbb{Z}_{q}\setminus\left\{ 0\right\} $.\par\end{flushleft}

\begin{flushleft}\textbf{Remark} \emph{If $\left\{ \left\{ 0\right\} ,E,F\right\} $
is convenient then $\forall A\in\mathcal{P}_{\widehat{\chi(E)}}{\cap\mathcal{P}}_{\widehat{\chi(F)}}$
one has $\left(\widehat{\chi(A)}\right)_{i}=\left(\widehat{\chi(A)}\right)_{j}$
if $i,j\in E$ or $i,j\in F$ and so if $\mathcal{P}_{\widehat{\chi(A)}}=\left\{ B_{1},\cdots,B_{r}\right\} $
then $E\subset B_{s}$ and $F\subset B_{t}$ for some $s$ and $t$}. \par\end{flushleft}

\paragraph*{Proof of the result \ref{sub:q-squareprime_R7}}

We consider below five possible cases of couples $(E,E')$ such that
$\left\{ \left\{ 0\right\} E,E'\right\} $ is convenient. In each
case we are going to apply the remark above with a suitable choice
of the set $A$, as well as Lemma 5.

\begin{enumerate}
\item $E=iH,\ p\nmid i,\ H<\mathbb{Z}_{q}^{\star}$ , $E'=jK,\ p\nmid j,\ K<\mathbb{Z}_{q}^{\star}$,
$H$ and $K$ are both of type 2. Using Eq. \ref{eq:lm4_3}, $\mathcal{P}_{\widehat{\chi(E)}}=\left\{ kH\vert\ k\in\mathbb{Z}_{q}\right\} $
and $\mathcal{P}_{\widehat{\chi(F)}}=\left\{ kH\vert\ k\in\mathbb{Z}_{q}\right\} $
so $\mathcal{P}_{\widehat{\chi(E)}}\cap\mathcal{P}_{\widehat{\chi(F)}}=\left\{ k(H\cap K)\vert\ k\in\mathbb{Z}_{q}\right\} $.
Taking $A=H\cap K$, $\mathcal{P}_{\widehat{\chi(A)}}=\left\{ l(H\cap K)\vert\ l\in\mathbb{Z}_{q}\right\} $,
one gets that $E\subset l(H\cap K)$ and $E'\subset m(H\cap K)$ for
some $l$ and $m$ which is possible only if $H=K$.
\item $E=iH,\ p\nmid i,\ H<\mathbb{Z}_{q}^{\star}$ , $H$ is of type 2
and $E'=jP^{-1}(L),\ L<\mathbb{Z}_{p}^{\star},\ p\nmid j$, $\mathcal{P}_{\widehat{\chi(F)}}=\left\{ \left\{ 0\right\} ,\mathbb{Z}_{q}^{\star},\left\{ pkL\vert\ p\nmid k\right\} \right\} $.
Taking $A=H\in\mathcal{P}_{\widehat{\chi(E)}}{\cap\mathcal{P}}_{\widehat{\chi(F)}}$
, $jP^{-1}(L)\subset mH$ for some $m$ which is in contradiction
with the fact that $H$ is of type 2.
\item $E=iH,\ p\nmid i,\ H<\mathbb{Z}_{q}^{\star}$ , $H$ is of type 2
and $E'=jpK,\ K<\mathbb{Z}_{q}^{\star}$, $\mathcal{P}_{\widehat{\chi(F)}}=\left\{ p\mathbb{Z}_{p},\left\{ tP^{-1}(K)\vert\ p\nmid t\right\} \right\} $.
Taking $A=H\cap P^{-1}(K)$ which a subgroup of type 2, $\mathcal{P}_{\widehat{\chi(A)}}=\left\{ s(H\cap P^{-1}(K))\vert\ s\in\mathbb{Z}_{q}\right\} $
and therefore $H\subset s(H\cap P^{-1}(K))$ and $pK\subset pt(H\cap P^{-1}(K))$
for some $s$ and $t$. We deduce that $pH=pK$.
\item $E=ipK,\ p\nmid i,\ K<\mathbb{Z}_{q}^{\star}$ and $E'=jpL,\ p\nmid j,\ L<\mathbb{Z}_{q}^{\star}$,
$\mathcal{P}_{\widehat{\chi(E)}}{\cap\mathcal{P}}_{\widehat{\chi(F})}=\left\{ p\mathbb{Z}_{p},\left\{ mP^{-1}(K\cap L)\vert\ p\nmid m\right\} \right\} $.
Taking now $A=P^{-1}(K\cap L)$, $\mathcal{P}_{\widehat{\chi(A)}}=\left\{ \left\{ 0\right\} ,\mathbb{Z}_{q}^{\star},\left\{ pt(K\cap L)\vert\ p\nmid t\right\} \right\} $
yielding $K=L$.
\item $E=iP^{-1}(K),\ K<\mathbb{Z}_{p}^{\star},\ p\nmid i$ and $E'=jP^{-1}(L),\ L<\mathbb{Z}_{p}^{\star},\ p\nmid j$.
Taking $A=p(K\cap L)\in\mathcal{P}_{\widehat{\chi(E)}}{\cap\mathcal{P}}_{\widehat{\chi(F)}}=\left\{ \left\{ 0\right\} ,\mathbb{Z}_{q}^{\star},\left\{ pt(K\cap L)\vert\ p\nmid t\right\} \right\} $,
$\mathcal{P}_{\widehat{\chi(A)}}=\left\{ p\mathbb{Z}_{p},\left\{ mP^{-1}(K\cap L)\vert\ p\nmid m\right\} \right\} $
one gets $K=L$.
\end{enumerate}
We now take $\mathcal{E}=\left\{ E_{1},\cdots,E_{r}\right\} $ and
$\mathcal{F}=\left\{ \mathcal{P}_{\widehat{\chi\left(aP{}^{-1}(H)\right)}}=\left\{ \left\{ 0\right\} ,\mathbb{Z}_{q}^{\star},\left\{ pmH\vert\ p\nmid m\right\} \right\} F_{1},\cdots,F_{r}\right\} $
verifying Eq. \ref{eq:1}. If one of the $E_{i}$'s is of the type
of Lemma 6, then $\mathcal{E}=\left\{ \left\{ 0\right\} ,\mathbb{Z}_{q}\setminus\left\{ 0\right\} \right\} $.
From now on we consider the case where no $E_{k}$ is of this type.
So all the $E_{k}$'s are either $iH$, or $jpK$ or $\left\{ 0\right\} $,
with $H<\mathbb{Z}_{q}^{\star}$$\ p\nmid i$ or $K<\mathbb{Z}_{p}^{\star}$$\ p\nmid j$.
It is clear that if $E,\ E'$ (distinct) are in $\mathcal{E}$, then
$\left\{ \left\{ 0\right\} E,E'\right\} $ is convenient. Suppose
there is a set $E_{k}=iH$ with $H<\mathbb{Z}_{q}^{\star}$$\ p\nmid i$
with $H$ of type 2, then using the three first points above one gets
that the $E_{l}$ are the class modulo $H$. Using now the last two
points, we see that $\forall E\in\mathcal{E}\setminus\left\{ 0\right\} \ $
either $E=iP^{-1}(K),\ p\nmid i$ or $E=jpL,\ p\nmid j$. Thus we
have shown that the only possible $P$-stable patterns are those occurring
in the paragraph \ref{sub:q-squareprime_R7}.

We now verify that these cases are \emph{indeed} $P$-stable. Obviously$\left\{ \left\{ 0\right\} ,\mathbb{Z}_{q}\setminus\left\{ 0\right\} \right\} $and
$\left\{ \alpha L\right\} $ are $P$-stable. Let us now take $\mathcal{E}=\left\{ \left\{ 0\right\} ,aP^{-1}(H),bpK\right\} $
with $H,K<\mathbb{Z}_{p}^{\star}$ and $a,b\in\mathbb{Z}_{p}^{\star}$,
$\mathcal{P}_{\widehat{\chi(aP{}^{-1}(H)}}=\left\{ \left\{ 0\right\} ,\mathbb{Z}_{q}^{\star},\left\{ pmH\vert\ p\nmid m\right\} \right\} $
and $\mathcal{P}_{\widehat{\chi(bpK)}}=\left\{ p\mathbb{Z}_{p},\left\{ tP^{-1}K\vert\ p\nmid t\right\} \right\} $.
Note that these two sets are independent of $a$ and $b$. Showing
that $\mathcal{E}$ is product stable amounts to showing that it is
convenient. The only possible $A$ in $\mathcal{P}_{\widehat{\chi\left(P{}^{-1}(H)\right)}}\cap\mathcal{P}_{\widehat{\chi(pK)}}$
are $\left\{ 0\right\} $, $pmH$ and $tP^{-1}(K)$ with $p\nmid m$
and $p\nmid t$. When $A=pmH$ then $\mathcal{P}_{\widehat{\chi(A)}}=\left\{ p\mathbb{Z}_{p},\left\{ mP^{-1}(H)\vert\ p\nmid m\right\} \right\} $
while when $A=tP^{-1}(K)$$,\mathcal{P}_{\widehat{\chi(A)}}=\left\{ \left\{ 0\right\} ,\mathbb{Z}_{q}^{\star},\left\{ pmHK\vert\ p\nmid m\right\} \right\} $
. In both cases Eq. \ref{eq:1} is verified which completes the proof
of \ref{sub:q-squareprime_R7}.

We are now in position to compute the number of stable patterns. We
recall that a cyclic group of order $n$ has $\tau(n)$ subgroups
so that $\mathbb{Z}_{q}^{\star}$ has $\tau(p^{2}-p)=2\tau(p-1)$
subgroups, but there are as many subgroups of type 1 as subgroups
in $\mathbb{Z}_{p}^{\star}$ $i.e.$ $\tau(p-1)$, so there are $\tau(p-1)$
subgroups of type 2. Finally $\left|\left\{ \left\{ 0\right\} ,aP^{-1}(H),bpK\right\} \right|=\left(\tau(p-1)\right)^{2}$,
yielding the stated result. 

In appendix we give as an illustration the example of all stable patterns
for $q=25=5^{2}$.

\subsubsection{Proof of \ref{sub:inte-fam_R8}: an integrable one-parameter family
of integrable patterns}

\begin{itemize}
\item Clearly for the pattern defined by Eq \ref{eq:int}. \[
\mbox{card}(E_{1})=\mbox{card}(E_{2})=\frac{q-1}{2}\]

\item If $k\in E_{1}$then there exists $a$ such that $k=a^{2}\mbox{ mod }q$
therefore \[
kE_{1}=a^{2}\left\{ i^{2}\mbox{ mod }q\right\} =E_{1}\]
 so $\sum_{i\in E_{1}}w^{ki}$ is independent of $k$ $\forall k\in E_{1}$.
Let us define $A$ and $A'$\[
A\equiv\sum_{i\in E_{1}}w^{k_{1}i},\ \ A'\equiv\sum_{i\in E_{2}}w^{k_{2}i}\]
where $k_{1}\in E_{1}$ and $k_{2}\in E_{2}$. One has $A+A'+1=0$. 
\end{itemize}
Let us recall the Gauss's result \cite{GaussEquality} \[
\sum_{j=0}^{q-1}\exp\frac{2\imath\pi j^{2}}{q}=\left\{ \begin{array}{ccc}
(1+\imath)\sqrt{q} &  & q=0\mbox{ mod }4\\
\sqrt{q} &  & q=1\mbox{ mod }4\\
0 &  & q=2\mbox{ mod }4\\
\imath\sqrt{q} &  & q=3\mbox{ mod }4\end{array}\right.\]
from which one deduces\[
A=\frac{\epsilon_{q}\sqrt{q}-1}{2}\]
with\[
\epsilon_{q}=\left\{ \begin{array}{ccc}
1 & \mbox{ if } & q\equiv1\mbox{ mod }4\\
\imath & \mbox{ if } & q\equiv3\mbox{ mod }4\end{array}\right.\]

\begin{itemize}
\item Finally, the Fourier transform is represented by the $3\times3$ matrix:\begin{equation}
F_{q}=\left(\begin{array}{ccc}
1 & \frac{q-1}{2} & \frac{q-1}{2}\\
1 & \frac{\epsilon_{q}\sqrt{q}-1}{2} & \frac{-\epsilon_{q}\sqrt{q}-1}{2}\\
1 & \frac{-\epsilon_{q}\sqrt{q}-1}{2} & \frac{\epsilon_{q}\sqrt{q}-1}{2}\end{array}\right)\label{eq:fou}\end{equation}

\end{itemize}
When $q=1\mbox{ mod }4$ then $\epsilon_{q}=1$, Eq. \ref{eq:fou}
corresponds to Eq17 and Eq 18 of \cite{defprmationdynamicschiralPotts}.
We deduce that the integrable mapping discovered in this reference
corresponds to a spin edge model of lattice statistical mechanics
when $q$ is a prime number with $q=1\mbox{ mod }4$, the pattern
is explicitly defined \emph{}by Eq. \ref{eq:int}. The homogeneous
expression of the transformation $K:\;(x,y,z)\rightarrow(X,Y,Z)$
can then easily be found. If one introduces then the inhomogeneous
variables $u=\frac{y}{x}$ and $v=\frac{z}{x}$ a $K-$invariant having
a particularly simple form is \[
\Delta=\frac{(u-v)^{2}}{(2uv-u-v)(u+v-2)}\left(q+2\frac{uv-u-v+1}{u+v}\right)\]

When $q=-1\mbox{ mod }4$ then $\epsilon_{q}=\imath$ and the corresponding
mapping is \emph{not} integrable. However a complexity reduction occurs
and using the method developed in \cite{calculseminumcomplexite,OnTheComp}
one finds for the generating function of the degree defined in Eq.
\ref{eq:genfun}\[
f(x)=\frac{1}{(1-x)(1-x-x^{2})}\]
leading to a complexity\[
\lambda=\frac{\sqrt{5}-1}{2}\simeq1.618034\]

\subsubsection{Mapping of the six families of stable patterns of \ref{sub:3coul-genera_R9}}

These results are verified by direct inspection. Using the methodology
and notation of \cite{calculseminumcomplexite}, the collineation
$C$, the inverse $I$, the generic (i.e. arbitrary $q$) degree generating
function $f$ and therefore the complexity can be computed for $P_{1}$,
$Q_{1}$, $P_{2}$ and $Q_{2}$ . 

\begin{itemize}
\item for $P_{1}$ and $Q_{1}$ the mapping are trivial. $P_{1}$ leads
to a linear mapping with generating function $\frac{1}{1-x}$, and
$Q_{1}$ to a  mapping ($K_{Q_{2}}^{2}=-1$).\[
C_{P_{1}}=\left(\begin{array}{cc}
1 & 1\\
1 & 1-q\end{array}\right),\quad I_{P_{1}}\left(\begin{array}{c}
x_{0}\\
x_{1}\end{array}\right)=\left(\begin{array}{c}
x_{0}+(2-q)x_{1}\\
-x_{2}\end{array}\right)\]
\[
C_{Q_{1}}=\left(\begin{array}{cc}
1 & 1\\
1 & \epsilon_{q}\end{array}\right),\quad I_{Q_{1}}\left(\begin{array}{c}
x_{0}\\
x_{1}\end{array}\right)=\left(\begin{array}{c}
x_{1}\\
-x_{2}\end{array}\right)\]

\item for $P_{2}$ and $Q_{2}$, one can also find explicitly the expression
of the collineation and of the inverse. The various expressions depend
of the parity of $\frac{q}{2}$ (as mentioned $q$ should be even)
yielding four mappings all having the same degree generating function\[
f(x)=\frac{1+x}{(1-x)(1-2x)}\]
giving an integer complexity $\lambda=2$. Introducing $\epsilon_{q}=(-1)^{\frac{q}{2}}$
the corresponding collineation and inverse are given below \[
C_{P_{2}}=\left(\begin{array}{ccc}
1 & 2 & 1\\
\frac{1+\epsilon_{q}}{2} & \frac{1-\epsilon_{q}}{2} & -1\\
1 & -(q-2) & 1\end{array}\right),\quad I_{P_{2}}\left(\begin{array}{c}
x_{0}\\
x_{1}\\
x_{2}\end{array}\right)=\left(\begin{array}{c}
x_{0}^{2}-(q-2)x_{1}^{2}-(q-4)x_{0}x_{1}+\epsilon_{q}x_{0}x_{2}\\
-(x_{0}-\epsilon_{q}x_{2})x_{1}\\
-\epsilon_{q}x_{2}^{2}+\epsilon_{q}(q-2)x_{1}^{2}+(q-4)x_{2}x_{1}-x_{0}x_{2}\end{array}\right)\]
\[
C_{Q_{2}}^{\mbox{even}}=\left(\begin{array}{ccc}
\frac{q}{2}-1 & \frac{q}{2} & 1\\
1 & 0 & -1\\
\frac{q}{2}-1 & -\frac{q}{2} & 1\end{array}\right),\quad I_{Q_{2}}^{\mbox{even}}\left(\begin{array}{c}
x_{0}\\
x_{1}\\
x_{2}\end{array}\right)=\left(\begin{array}{c}
(2q-4)x_{0}^{2}-2qx_{1}^{2}+4x_{0}x_{2}\\
-4(x_{0}-x_{2})x_{1}\\
-(q-2)(q-4)x_{0}^{2}+q(q-2)x_{1}^{2}-4x_{2}^{2}-4(q-3)x_{0}x_{2}\end{array}\right)\]
 \[
C_{Q_{2}}^{\mbox{odd}}=\left(\begin{array}{ccc}
\frac{q}{2} & \frac{q}{2}-1 & 1\\
0 & 1 & -1\\
\frac{q}{2} & -\frac{q}{2}+1 & -1\end{array}\right),\quad I_{Q_{2}}^{\mbox{odd}}\left(\begin{array}{c}
x_{0}\\
x_{1}\\
x_{2}\end{array}\right)=\left(\begin{array}{c}
4(x_{1}-x_{2})x_{0}\\
2qx_{0}^{2}+(4-2q)x_{1}^{2}-4x_{1}x_{2}\\
-q(q-2)x_{0}^{2}+(q-2)(q-4)x_{1}^{2}+4x_{2}^{2}+4(q-3)x_{1}x_{2}\end{array}\right)\]
The superscript refers to the parity of $\frac{q}{2}$.
\end{itemize}

%
\begin{comment}
There is a conjecture ref stating that there there is no cyclic Hadamard
matrices for $q>4$. This has been proved for $q<4\ 10^{11}$, with
possible exceptions of $q=108900,\ 47436840000,\ 26899280100$.
\end{comment}
{}

\section{Numerical results\label{sec:Numerical-results} }

\subsection{Computer-aided method}

To study the case where $q$ is neither a prime nor the square of
a prime, we use a computer. In principle there is no difficulty since
it is {}``only'' a matter of generating the patterns or the signed-patterns
and check the stability. To test the stability we can use the formula
Eq. \ref{eq:1}-\ref{eq:4}. However, in practice, this would be tractable
only for a very small value of $q$ since the number of patterns grows
extremely fast with $q$. 

If a partition $\mathcal{E}=\left\{ E_{1},\cdots,E_{r}\right\} $
is product stable then for any $k$, $E_{k}$ must be such that $i,j\in E_{k}\Rightarrow\left(\mbox{Cy}(\chi(E_{k}))\right)_{i}=\left(\mbox{Cy}(\chi(E_{k}))\right)_{j}$.
Since the cyclic matrices associated to a stable pattern form an algebra,
we see that any subset $E_{k}$ must be convenient (see Eq. \ref{eq:convenient}).
This simple remark tells that it is not necessary to generate \emph{all}
possible subset $E_{k}$, but only convenient sets: one can first
enumerate the $2^{q}$ subsets $E$ of $\left\{ 0,\cdots,q-1\right\} $and
keep only those which are convenient. The key point here is that a
subset $E$ is convenient irrespective of the way the remaining indices
are grouped into others subsets. Then, using a tree structure, we
can associate convenient subsets to make stable patterns. In order
to minimize the tree structure, one can also use condition on pairs
(or more) of index subsets, also deduced from Eq. \ref{eq:convenient}.
This procedure applies \emph{mutatis mutandis} to inverse-stability
when the matrices $\chi(E)$ are invertible. Note that this procedure
can also be applied to the search of non cyclic stable patterns. 

In the case of cyclic matrix we can go further and avoid the consideration
of all possible subsets, retaining only the convenient sets. Indeed
using the result of section \ref{sub:Admissible-subsets_R3} one can
generate \emph{directly} the admissible sets. We then consider these
sets as {}``atoms'' to be combined to produce the patterns. This
can also be implemented using a tree structure. In the results shown
below, we did not use this last remark.

\subsection{Results}

In table \ref{tab:Number-of-P-} we present the number of $P$-stable
and $I\bar{P}$-stable patterns and signed-patterns for $2\le q\le8$.
In table \ref{tab:Number-of-product-stable} we present the number
of $P$-stable patterns for larger values of $q$. All these numbers
have been found using the algorithm presented above. We have verified
that for $q$ prime (i.e. $q=2,\ 3,\ 5,\ 7,\ 11,\ 13,\ 17,\ 19,\ 23,\ 29$)
it corresponds to the result of section \ref{sub:q-prime_R6}, and
for $q$ the square of a prime ($q=4,\ 9,\ 25$) to the result of
section \ref{sub:q-squareprime_R7}. In Table \ref{tab:Number-of-P-}
the first line is the number of convenient sets. The explicit expression
of each pattern is not given in the text, but can be downloaded from
the site \cite{download}. However the case $q=8$ is given in detail
in Appendix \ref{app:q8}. Finally a maple program to generate all
the stable patterns for $q=p^{2}$ with $p$ prime can be download
from \cite{download}.

\begin{table}
\begin{tabular}{|c|c|c|c|c|c|c|c|}
\hline 
$q$&
2&
3&
4&
5&
6&
7&
8\tabularnewline
\hline
\hline 
\# tested pattern&
3&
11&
49&
257&
1539&
10299&
75905\tabularnewline
\hline
\hline 
$P$-Stable Pattern&
1&
2&
3&
3&
7&
4&
10\tabularnewline
\hline 
$P$-Stable signed-Pattern&
0&
0&
6&
0&
3&
0&
17\tabularnewline
\hline 
$I\bar{P}$-Stable Pattern&
0&
0 &
2&
0&
3&
0&
11\tabularnewline
\hline 
$I\bar{P}$-Stable signed-Pattern&
0&
0&
2&
0&
1&
0&
19\tabularnewline
\hline
\hline 
Total&
1&
2&
13&
3&
14&
4&
57\tabularnewline
\hline
\end{tabular}

\caption{Number of $P$- and $I\bar{P}$-stable patterns and signed-patterns.\label{tab:Number-of-P-}}

\begin{centering}\begin{tabular}{|c|c|c|c|c|c|c|c|c|c|}
\hline 
$q$&
9&
10&
11&
12&
13&
14&
15&
16&
17\tabularnewline
\hline 
\# stable patterns&
7&
10&
4&
32&
6&
13&
21&
37&
5\tabularnewline
\hline
\hline 
$q$&
18&
19&
20&
21&
22&
23&
24&
25&
26\tabularnewline
\hline 
\# stable patterns&
42&
6&
47&
28&
14&
5&
172&
13&
19\tabularnewline
\hline
\hline 
$q$&
27&
28&
29&
30&
31&
37&
41&
43&
49\tabularnewline
\hline 
\# stable patterns&
25&
61&
7&
148&
8&
9&
8&
8&
21\tabularnewline
\hline
\end{tabular}\par\end{centering}

\caption{Number of product-stable patterns.\label{tab:Number-of-product-stable}}
\end{table}

\section{Conclusion}

In this paper we have shown several results concerning stable patterns
in the case of cyclic matrices. The notion of signed-pattern arises
naturally when one studies $I\bar{P}$-stability as a consequence
of a duality between cyclic matrices and their Fourier transform.
We find in particular an exact correspondence between $I\bar{P}-$stable
patterns and $P-$stable signed-patterns, which justifies, \emph{a
posteriori}, the introduction of signed-patterns in this cyclic matrix
context. The main results Eq. \ref{eq:1}-Eq. \ref{eq:4} enable to
find all $I$-stable patterns and signed-patterns when the number
of states is a prime or the square of a prime, and to find some, but
not all, stable patterns for composite integer values of $q$. \textcolor{black}{This
provides examples of birational transformations of an arbitrary large
number of variables.} We have computed the complexity of the corresponding
transformations in some cases, finding a complexity reduction. In
particular we have recovered a one-parameter family of integrable
transformations, for which we have given explicitly the matrix representation
when it exists. The case of the monocolor $I$-stable signed-patterns
has been solved, demonstrating a conjecture about Hadamard matrices
in a particular case. We also present an algorithm to find $I$-stable
patterns. Although this algorithm is exponential, it can be used for
not too large values of $q$.

It would be interesting to generalize our result to arbitrary value
of $q$ and to perform a more systematic analysis of the complexity
of the associated birational transformation. Finally the same problem
for non cyclic matrices should also be investigated, but it seems
to us that it becomes a very complicated task, as a consequence of
the loss of the discrete Fourier transform. 

\appendix

\section{Stable patterns for $q=8$\label{app:q8}}

We list below the stable patterns for $q=8$. The number before the
eight letters between bracket is the arbitrary label of the pattern.
The sequence of eight letters designates the first row of the cyclic
matrix in the direct space, and the diagonal of the matrix in the
Fourier space. When a letter is repeated (resp. negated) this means
that the two corresponding entries of the matrix are equal, (resp.
opposite). For example the pattern number 10 below corresponds to
\[
M_{\mbox{Fourier}}^{10}=\left(\begin{array}{cccccccc}
a & 0 & 0 & 0 & 0 & 0 & 0 & 0\\
0 & b & 0 & 0 & 0 & 0 & 0 & 0\\
0 & 0 & -b & 0 & 0 & 0 & 0 & 0\\
0 & 0 & 0 & b & 0 & 0 & 0 & 0\\
0 & 0 & 0 & 0 & c & 0 & 0 & 0\\
0 & 0 & 0 & 0 & 0 & b & 0 & 0\\
0 & 0 & 0 & 0 & 0 & 0 & -b & 0\\
0 & 0 & 0 & 0 & 0 & 0 & 0 & b\end{array}\right),\qquad M_{\mbox{direct}}^{10}=\left(\begin{array}{cccccccc}
a & b & -b & b & c & b & -b & b\\
b & a & b & -b & b & c & b & -b\\
-b & b & a & b & -b & b & c & b\\
b & -b & b & a & b & -b & b & c\\
c & b & -b & b & a & b & -b & b\\
b & c & b & -b & b & a & b & -b\\
-b & b & c & b & -b & b & a & b\\
b & -b & b & c & b & -b & b & a\end{array}\right)\]
and the cyclic matrix associated to pattern 17 with first row $[a,b,a,b,c,b,a,b]$
becomes by fourier transform the diagonal matrix $M_{\mbox{fourier}}^{10}$
to which one associates in the direct space the cyclic matrix $M_{\mbox{direct}}^{10}$
(see above).

\begin{center}\begin{eqnarray*}
\mbox{ Direct space}\qquad\rightarrow\mbox{Fourier space} &  & \mbox{Direct space}\qquad\rightarrow\mbox{Fourier space}\\
P\mbox{-stable pattern}\rightarrow P\mbox{-stable pattern} &  & I\bar{P}\mbox{-stable pattern}\rightarrow P\mbox{-stable signed-pattern}\\
9[a,b,c,d,e,b,f,d]\rightarrow9[a,b,c,d,e,b,f,d] &  & 4[a,a,a,a,b,a,a,a]\rightarrow55[a,b,-b,b,-b,b,-b,b]\\
15[a,b,c,b,d,b,c,b]\rightarrow15[a,b,c,b,d,b,c,b] &  & 6[a,b,c,b,a,d,e,d]\rightarrow35[a,b,c,d,e,-d,c,-b]\\
22[a,b,b,b,b,b,b,b]\rightarrow22[a,b,b,b,b,b,b,b] &  & 11[a,b,a,c,a,d,a,e]\rightarrow31[a,b,c,d,e,-b,-c,-d]\\
39[a,b,c,d,e,d,c,b]\rightarrow39[a,b,c,d,e,d,c,b] &  & 14[a,b,c,d,a,b,e,d]\rightarrow21[a,b,c,-b,d,b,e,-b]\\
48[a,b,c,d,e,f,g,h]\rightarrow48[a,b,c,d,e,f,g,h] &  & 17[a,b,a,b,c,b,a,b]\rightarrow10[a,b,-b,b,c,b,-b,b]\\
51[a,b,c,b,d,e,c,e]\rightarrow51[a,b,c,b,d,e,c,e] &  & 18[a,b,c,b,a,d,c,d]\rightarrow5[a,b,c,b,d,-b,c,-b]\\
3[a,b,c,b,d,b,e,b]\leftrightarrow53[a,b,c,d,e,b,c,d] &  & 20[a,b,c,d,a,d,c,b]\rightarrow47[a,b,c,-b,d,-b,c,b]\\
24[a,b,c,b,c,b,c,b]\leftrightarrow56[a,b,b,b,c,b,b,b] &  & 25[a,a,b,a,a,a,c,a]\rightarrow37[a,b,c,-b,-c,b,c,-b]\\
 &  & 29[a,b,c,d,a,e,c,f]\rightarrow13[a,b,c,d,e,-b,f,-d]\\
 &  & 34[a,b,c,b,a,b,d,b]\rightarrow49[a,b,c,-b,d,b,c,-b]\\
 &  & 54[a,b,c,d,a,d,e,b]\rightarrow44[a,b,c,-b,d,e,c,-e]\end{eqnarray*}
\par\end{center}

\begin{center}\begin{eqnarray*}
\mbox{Direct space}\qquad\rightarrow\mbox{Fourier space} &  & \mbox{Direct space}\qquad\qquad\rightarrow\;\mbox{Fourier space}\\
P\mbox{-stable signed-pattern}\rightarrow I\bar{P}\mbox{-stable pattern} &  & I\bar{P}\mbox{-stable signed-pattern}\rightarrow I\bar{P}\mbox{-stable signed-pattern}\\
5[a,b,c,b,d,-b,c,-b]\rightarrow18[a,b,c,b,a,d,c,d] &  & 8[a,b,c,-b,a,-b,c,b]\rightarrow8[a,b,c,-b,a,-b,c,b]\\
10[a,b,-b,b,c,b,-b,b]\rightarrow17[a,b,a,b,c,b,a,b] &  & 16[a,-a,a,-a,b,-a,a,-a]\rightarrow16[a,-a,a,-a,b,-a,a,-a]\\
13[a,b,c,d,e,-b,f,-d]\rightarrow29[a,b,c,d,a,e,c,f] &  & 19[a,b,c,d,-a,b,-c,d]\rightarrow19[a,b,c,d,-a,b,-c,d]\\
21[a,b,c,-b,d,b,e,-b]\rightarrow14[a,b,c,d,a,b,e,d] &  & 23[a,b,c,d,a,-d,e,-b]\rightarrow23[a,b,c,d,a,-d,e,-b]\\
31[a,b,c,d,e,-b,-c,-d]\rightarrow11[a,b,a,c,a,d,a,e] &  & 27[a,b,c,b,a,-b,c,-b]\rightarrow27[a,b,c,b,a,-b,c,-b]\\
35[a,b,c,d,e,-d,c,-b]\rightarrow6[a,b,c,b,a,d,e,d] &  & 32[a,b,c,d,-a,e,-c,f]\rightarrow32[a,b,c,d,-a,e,-c,f]\\
37[a,b,c,-b,-c,b,c,-b]\rightarrow25[a,a,b,a,a,a,c,a] &  & 38[a,b,c,d,a,-b,c,-d]\rightarrow38[a,b,c,d,a,-b,c,-d]\\
44[a,b,c,-b,d,e,c,-e]\rightarrow54[a,b,c,d,a,d,e,b] &  & 42[a,b,c,-b,a,d,e,-d]\rightarrow42[a,b,c,-b,a,d,e,-d]\\
47[a,b,c,-b,d,-b,c,b]\rightarrow20[a,b,c,d,a,d,c,b] &  & 50[a,b,c,-b,a,b,d,-b]\rightarrow50[a,b,c,-b,a,b,d,-b]\\
49[a,b,c,-b,d,b,c,-b]\rightarrow34[a,b,c,b,a,b,d,b] &  & 1[a,b,c,b,a,-b,-c,-b]\leftrightarrow52[a,b,a,c,a,-c,a,-b]\\
55[a,b,-b,b,-b,b,-b,b]\rightarrow4[a,a,a,a,b,a,a,a] &  & 2[a,b,c,-b,-a,-b,c,b]\leftrightarrow45[a,b,-a,c,a,c,-a,b]\\
 &  & 7[a,b,c,d,a,-d,c,-b]\leftrightarrow36[a,b,c,b,a,-b,d,-b]\\
 &  & 12[a,b,c,-b,a,-b,-c,b]\leftrightarrow41[a,b,a,-b,a,c,a,-c]\\
 &  & 26[a,b,-a,b,a,c,-a,c]\leftrightarrow57[a,b,c,b,-a,-b,c,-b]\\
 &  & 28[a,b,c,-b,a,-b,d,b]\leftrightarrow33[a,b,c,-b,a,d,c,-d]\\
 &  & 30[a,-a,b,-a,a,-a,c,-a]\leftrightarrow40[a,b,-a,-b,c,b,-a,-b]\\
 &  & 43[a,b,-a,c,a,d,-a,e]\leftrightarrow46[a,b,c,d,-a,-b,e,-d]\end{eqnarray*}
\par\end{center}

In the following we show in detail that Eq. \ref{eq:4} is indeed
verified on the example of the Fourier related pair of signed-patterns
labeled 7 and 36 above. We used the letter $E$ for pattern 7 and
$F$ for pattern 36, and $\omega=\exp\frac{2\pi}{8}\imath$.\[
\chi(E_{0}^{+})=\left(\begin{array}{c}
1\\
0\\
0\\
0\\
1\\
0\\
0\\
0\end{array}\right),\chi(E_{1}^{+})=\left(\begin{array}{c}
0\\
1\\
0\\
0\\
0\\
0\\
0\\
0\end{array}\right),\chi(E_{1}^{-})=\left(\begin{array}{c}
0\\
0\\
0\\
0\\
0\\
0\\
0\\
1\end{array}\right),\chi(E_{2}^{+})=\left(\begin{array}{c}
0\\
0\\
1\\
0\\
0\\
0\\
1\\
0\end{array}\right),\chi(E_{3}^{+})=\left(\begin{array}{c}
0\\
0\\
0\\
1\\
0\\
0\\
0\\
0\end{array}\right),\chi(E_{3}^{-})=\left(\begin{array}{c}
0\\
0\\
0\\
0\\
0\\
1\\
0\\
0\end{array}\right)\]
\[
\widehat{\chi(E_{0}^{+})}=\left(\begin{array}{c}
2\\
0\\
2\\
0\\
2\\
0\\
2\\
0\end{array}\right),\widehat{\chi(E_{1}^{+})}=\left(\begin{array}{c}
1\\
\omega\\
\imath\\
-\bar{\omega}\\
-1\\
-\omega\\
-\imath\\
\bar{\omega}\end{array}\right),\widehat{\chi(E_{1}^{-})}=\left(\begin{array}{c}
1\\
\bar{\omega}\\
-\imath\\
-\omega\\
-1\\
-\bar{\omega}\\
\imath\\
\omega\end{array}\right),\widehat{\chi(E_{2}^{+})}=\left(\begin{array}{c}
2\\
0\\
-2\\
0\\
2\\
0\\
-2\\
0\end{array}\right),\widehat{\chi(E_{3}^{+})}=\left(\begin{array}{c}
1\\
-\bar{\omega}\\
-\imath\\
\omega\\
-1\\
\bar{\omega}\\
\imath\\
-\omega\end{array}\right),\widehat{\chi(E_{3}^{-})}=\left(\begin{array}{c}
1\\
-\omega\\
\imath\\
\bar{\omega}\\
-1\\
\omega\\
-\imath\\
-\bar{\omega}\end{array}\right)\]
\begin{equation}
\widehat{\chi(E_{0}^{+})}=\left(\begin{array}{c}
2\\
0\\
2\\
0\\
2\\
0\\
2\\
0\end{array}\right),\widehat{\chi(E_{1}^{+})}-\widehat{\chi(E_{1}^{-})}=\left(\begin{array}{c}
0\\
\sqrt{2}\imath\\
2\imath\\
\sqrt{2}\imath\\
0\\
-\sqrt{2}\imath\\
-2\imath\\
-\sqrt{2}\imath\end{array}\right),\widehat{\chi(E_{2}^{+})}=\left(\begin{array}{c}
2\\
0\\
-2\\
0\\
2\\
0\\
-2\\
0\end{array}\right),\widehat{\chi(E_{3}^{+})}-\widehat{\chi(E_{3}^{-})}=\left(\begin{array}{c}
0\\
\sqrt{2}\imath\\
-2\imath\\
\sqrt{2}\imath\\
0\\
-\sqrt{2}\imath\\
2\imath\\
-\sqrt{2}\imath\end{array}\right)\label{eq:examE}\end{equation}
on another hand\begin{equation}
\chi(F_{0}^{+})=\left(\begin{array}{c}
1\\
0\\
0\\
0\\
1\\
0\\
0\\
0\end{array}\right),\chi(F_{1}^{+})-\chi(F_{1}^{-})=\left(\begin{array}{c}
0\\
1\\
0\\
1\\
0\\
-1\\
0\\
-1\end{array}\right),\chi(F_{2}^{+})=\left(\begin{array}{c}
0\\
0\\
1\\
0\\
0\\
0\\
0\\
0\end{array}\right),\chi(F_{3}^{+})=\left(\begin{array}{c}
0\\
0\\
0\\
0\\
0\\
0\\
1\\
0\end{array}\right)\label{eq:examF}\end{equation}
It is then straightforward to verify that the four vectors of Eq.
\ref{eq:examE} and Eq. \ref{eq:examF} span the subspace:\begin{eqnarray*}
\widehat{\chi(E_{0}^{+})} & = & 2\chi(F_{0}^{+})+2\chi(F_{2}^{+})+2\chi(F_{3}^{+})\\
\widehat{\chi(E_{1}^{+})}-\widehat{\chi(E_{1}^{+})} & = & \sqrt{2}\imath\left(\chi(F_{1}^{+})-\chi(F_{1}^{-})\right)+2\imath\chi(F_{2}^{+})-2\imath\chi(F_{3}^{+})\\
\widehat{\chi(E_{2}^{+})} & = & 2\chi(F_{0}^{+})-2\chi(F_{2}^{+})-2\chi(F_{3}^{+})\\
\widehat{\chi(E_{3}^{+})}-\widehat{\chi(E_{3}^{-})} & = & \sqrt{2}\imath\left(\chi(F_{1}^{+})-\chi(F_{1}^{-})\right)-2\imath\chi(F_{2}^{+})+2\imath\chi(F_{3}^{+})\end{eqnarray*}

\section{Stable patterns for $q=25=5^{2}$}

We list, as an illustration, the stable patterns for $q=25.$ There
are $1+\tau(4)+\tau^{2}(4)=1+3+3^{2}=13$ such stable patterns.

Firstly there is the simple pattern corresponding to the standard
Potts model: \begin{eqnarray*}
1 &  & [a,b,b,b,b,b,b,b,b,b,b,b,b,b,b,b,b,b,b,b,b,b,b,b,b]\end{eqnarray*}
then the pattern corresponding to the subgroup of $\mathbb{Z}_{25}^{\star}=\left\{ 1,2,3,4,6,7,8,9,11,12,13\cdots,24\right\} $
:

\begin{center}\begin{eqnarray*}
L_{1} & = & \mathbb{Z}_{24}^{\star}\\
L_{2} & = & \left\{ 1,4,6,9,11,14,16,19,21,24\right\} \\
L_{3} & = & \left\{ 1,6,11,16,21\right\} \\
L_{4} & = & \left\{ 1,7,18,24\right\} \\
L_{5} & = & \left\{ 1,24\right\} \\
L_{6} & = & \left\{ 1\right\} \end{eqnarray*}
\par\end{center}

\begin{flushleft}which gives the six patterns\begin{eqnarray*}
2 &  & [a,b,b,b,b,c,b,b,b,b,c,b,b,b,b,c,b,b,b,b,c,b,b,b,b]\\
3 &  & [a,b,c,c,b,d,b,c,c,b,e,b,c,c,b,e,b,c,c,b,d,b,c,c,b]\\
4 &  & [a,b,c,d,e,f,b,c,d,e,g,b,c,d,e,h,b,c,d,e,i,b,c,d,e]\\
5 &  & [a,c,d,e,e,b,f,c,f,g,b,d,g,g,d,b,g,f,c,f,b,e,e,d,c]\\
6 &  & [a,d,e,f,g,b,h,i,j,k,c,l,m,m,l,c,k,j,i,h,b,g,f,e,d]\\
7 &  & [a,g,h,i,j,c,k,l,n,b,d,p,m,o,q,e,r,s,t,u,f,v,w,x,y]\end{eqnarray*}
Finally the remaining patterns are computed from$\mathbb{Z}_{5}^{\star}=\left\{ 1,2,3,4\right\} $:\begin{eqnarray*}
K_{1}=\left\{ 1,2,3,4\right\}  &  & P^{-1}(K_{1})=\mathbb{Z}_{25}^{\star}\\
K_{2}=\left\{ 1,4\right\}  &  & P^{-1}(K_{2})=\left\{ 1,4,6,9,11,14,16,19,21,24\right\} \\
K_{3}=\left\{ 1\right\}  &  & P^{-1}(K_{3})=\left\{ 1,6,11,16,21\right\} \end{eqnarray*}
yielding the six last patterns:\begin{eqnarray*}
8 &  & [a,b,b,b,b,c,b,b,b,b,d,b,b,b,b,d,b,b,b,b,c,b,b,b,b]\\
9 &  & [a,b,b,b,b,c,b,b,b,b,d,b,b,b,b,e,b,b,b,b,f,b,b,b,b]\\
10 &  & [a,c,d,d,c,b,c,d,d,c,b,c,d,d,c,b,c,d,d,c,b,c,d,d,c]\\
11 &  & [a,b,c,c,b,d,b,c,c,b,e,b,c,c,b,f,b,c,c,b,g,b,c,c,b]\\
12 &  & [a,c,d,e,f,b,c,d,e,f,b,c,d,e,f,b,c,d,e,f,b,c,d,e,f]\\
13 &  & [a,d,e,f,g,b,d,e,f,g,c,d,e,f,g,c,d,e,f,g,b,d,e,f,g]\end{eqnarray*}
\par\end{flushleft}


\begin{thebibliography}{10}
\bibitem{OnTheComp}\emph{On the complexity of some birational transformations}\\
J-C. Anglès d'Auriac, J-M. Maillard and C.M. Viallet.\\
J. Phys. \textbf{A}: Math. Gen. 39 3641-3654 (2006) 

\bibitem{biratfrommecastat}\emph{Integrable Coxeter groups} \\
M.P. Bellon, J-M. Maillard and C.M. Viallet,\\
Physics Letters. \textbf{A 159} 221-232 (1991)

\bibitem{BaBook} \emph{Exactly Solved Model in lattice statistical
Mechanics}.\\
 Baxter, R. J. \\
Academic Press, New. York. (1982)

\bibitem{MeAnMaRo94}\emph{Phase diagram of a six-state chiral Potts
model}\\
H.Meyer, J-C. Anglès d'Auriac, J-M. Maillard\\
Physica \textbf{A 208} 223-236 (1994)

\bibitem{definingcomplex1} \emph{Almost integrable mappings}\\
S. Boukraa, J-M. Maillard, and G. Rollet\emph{}\\
 \emph{}Int. J. Mod.Phys. \textbf{B8} 137-174 (1994\emph{)}

\bibitem{definingcomplex2}\emph{Integrable mappings and polynomial
growth}\\
S. Boukraa, J-M. Maillard, and G. Rollet,\\
Physica \textbf{A 209}, 162-222 (1994)

\bibitem{definingcomplex3}\emph{Determinental identities on integrable
mappings}\\
S. Boukraa, J-M. Maillard, and G. Rollet\\
Int. J. Mod. Phys. \textbf{B8} 2157-2201 (1994)

\bibitem{asssch_jaeger} \emph{Towards a classification of spin models
in terms of association schemes}.\\
Jaeger, F.\\
In Progress in algebraic combinatorics (Fukuoka, 1993), vol. 24. Math.
Soc. Japan, Tokyo, 197-225 (1996)

\bibitem{BoseMesner}\emph{On linear associative algebras corresponding
to association schemes of partially balanced designs}.\\
R.C. Bose D.M. Mesner\\
Ann. Math. Statist. \textbf{10} 21-38 (1959)

\bibitem{conjecthada}A conjecture probably due to R. Paley. See\\
 http://en.wikipedia.org/wiki/Hadamard\_matrix\#The\_Hadamard\_conjecture

\bibitem{Hadamardmatrices} \emph{Hadamard matrice}\\
http://www.research.att.com/\textasciitilde{}njas/hadamard

\bibitem{CyclotomicInteger} \emph{Cyclotomic Integers and Finite
Geometry}.\\
B. Schmidt \\
 J. Am. Math. Soc. \textbf{12} 929-952 (1999)

\bibitem{FWu}\emph{The Potts model}\\
F.Y. Wu\\
Rev. Mod. Phys. \textbf{54}, 235 - 268 (1982)

\bibitem{chiralpotts1}\emph{Commuting transfer matrices in the chiral
Potts models: Solution to the star triangle equations with genus >1}\\
H. Au-Yang, B.M. McCoy, J.H.H. Perk, S. Tang and Y.M. Lin,\\
 Phys. Letts. \textbf{A123}, 219 (1987)

\bibitem{chiralpotts2}\emph{New solutions of the star-triangle relations
for the chiral Potts model}\\
R. Baxter, H. Au-Yang and J.H.H. Perk \\
Phys. Lett. \textbf{A 128}, 138 (1988)

\bibitem{MarcuRittenberg}\emph{The global symmetries of spin systems
defined on abelian groups.}\\
M. Marcu V. Rittenberg\\
J. Math. Phys \textbf{22} (12) 2740-2752 (1981)\\
J. Math. Phys \textbf{22} (12) 2753-2758 (1981)

\bibitem{moreintegmap}\emph{More integrable birational mappings}.\\
 N. Abarenkova, J. C. Anglès d'Auriac and J.M. Maillard.\\
 Physica \textbf{A237}, 123 (1997)

\bibitem{inverserelation}\emph{A new calculation method for partition
functions in some lattice models}\\
Yu.G. Stroganov\\
Phys. Lett. \textbf{A 74} 116 (1979)

\bibitem{inversrelatJM}\emph{Symmetry relations in exactly soluble
models}\\
M. T. Jaekel and J-M. Maillard\\
J. Phys. \textbf{A 15} 1309-1325 (1982)\\
\emph{Inverse functional relations on the Potts model}\\
M. T. Jaekel and J-M. Maillard\\
J. Phys. \textbf{A 15} 2241-2257 (1982)\\
Inversion functional relations for lattice models\\
M. T. Jaekel and J-M. Maillard\\
J. Phys. \textbf{A 16}, 1975-1992 (1983)

\bibitem{ClassifAMV}\emph{A classification of four-state spin edge
Potts models}. \\
J-C. Anglès d'Auriac, J-M. Maillard and C.M. Viallet.\\
J.Phys.\textbf{\emph{A}}: Math. Gen.35 9251-9272 (2002)

\bibitem{bellonviallet}\emph{Algebraic Entropy}\\
M.P. Bellon and C.-M. Viallet\\
Comm. Math. Phys. 204, 425-427 (1999)

\bibitem{calculseminumcomplexite}\emph{Growth-complexity spectrum
of some discrete dynamical systems}.\\
N. Abarenkova, J-C. Anglès d'Auriac, S. Boukraa and J-M. Maillard.\\
Physica \textbf{D} 130, 27 (1999)

\bibitem{complexitereduc} \emph{Factorization properties of birational
mappings}\\
S. Boukraa and J-M. Maillard,\\
 Physica \textbf{A 220} 403-470 (1995)

\bibitem{bedfordkim}\emph{On the degree growth of birational mappings
in Higher Dimension.}\\
E. Bedford and K. Kim\\
Journ. Geom. Analysis \textbf{14,} 4 p567-596 (2004)

\bibitem{GaussEquality}\emph{A Classical Introduction to Modern Number
Theory}\\
K. Ireland and M. Rosen\\
 Second Edition, Springer-Verlag (1990)

\bibitem{defprmationdynamicschiralPotts}\emph{Deformations of dynamics
associated to the chiral Potts model}.\\
M.P. Bellon, J-M. Maillard, G. Rollet and C-M. Viallet\\
Int. J. Mod. Phys. \textbf{B6} 3575-3584 (1992)

\bibitem{AdABoMa99}\emph{Functional relations in lattice lattice
statistical mechanics, enumerative combinatorics and discrete dynamical
systems}\\
J-C. Anglès d'Auriac, S. Boukraa and J-M. Maillard\\
Annals of Combinatorics  \textbf{3}, 131-158 (1999)

\bibitem{download}http://perso.neel.cnrs.fr/jean-christian.angles-dauriac/ 

\bibitem{moebiusfunction}\emph{Introduction to Analytic Number Theory}\\
T.M. Apostol\\
Springer-Verlag Berlin Heidelberg (1998)
\end{thebibliography}
\end{document}